\documentclass[fleqn,usenatbib]{mnras}

\usepackage[T1]{fontenc}
\usepackage{ae,aecompl}

\usepackage{epsfig,graphicx}	
\usepackage{amsmath}	
\usepackage{amssymb}	
\usepackage{natbib}
\usepackage{bm}
\usepackage{comment}
\usepackage{dcolumn}
\usepackage{tabu}
\usepackage{datetime}
\settimeformat{hhmmsstime}
\usepackage{multicol}
\usepackage{pdflscape}
\title[Jet Ejections in V404 Cygni]{
Extreme Jet Ejections from the Black Hole X-ray Binary V404 Cygni}

\author[A.J. Tetarenko et al.]{A.J. Tetarenko,$^{1}$\thanks{E-mail: tetarenk@ualberta.ca}
G.R. Sivakoff,$^{1}$ 
J.C.A. Miller-Jones,$^{2}$ 
\newauthor
E.W. Rosolowsky,$^{1}$
G. Petitpas,$^{3}$
M. Gurwell,$^{3}$ 
J. Wouterloot,$^{4}$ 
R. Fender,$^{5}$
\newauthor
S. Heinz,$^{6}$ 
D. Maitra,$^{7}$
S.B. Markoff,$^{8}$ 
S. Migliari,$^{9,10}$
M.P. Rupen,$^{11,12}$
\newauthor
A.P. Rushton,$^{5,13}$ 
D.M. Russell,$^{14}$ 
T.D. Russell,$^{8,2}$
and C.L. Sarazin$^{15}$ 
\\
$^{1}$Department of Physics, University of Alberta, CCIS 4-181, Edmonton, AB T6G 2E1, Canada\\
$^{2}$International Centre for Radio Astronomy Research- Curtin University, GPO Box U1987, Perth, WA 6845, Australia\\
$^{3}$Harvard-Smithsonian Center for Astrophysics, Cambridge, MA 02138\\
$^{4}$East Asian Observatory, 660 North Aohoku Place, University Park, Hilo, Hawaii 96720, USA\\
$^{5}$Department of Physics, Astrophysics, University of Oxford, Keble Road, Oxford OX1 3RH, UK\\
$^{6}$Astronomy Department, University of Wisconsin-Madison, 475. N. Charter St., Madison, WI 53706, USA\\
$^{7}$Department of Physics and Astronomy, Wheaton College, Norton, MA 02766, USA\\
$^{8}$Anton Pannekoek Institute for Astronomy, University of Amsterdam, P.O. Box 94249, NL-1090 GE Amsterdam, the Netherlands\\
$^{9}$Department of Astronomy and Meteorology, 
University of Barcelona, Mart\'i i Franqu\`es 1, 08028 Barcelona, Spain\\
$^{10}$XMM-Newton Science Operations Centre, ESAC/ESA, PO Box 78, 28691 Villanueva de la Ca\~nada, Madrid, Spain\\
$^{11}$National Research Council, Herzberg Astronomy and Astrophysics, 717 White Lake Road, PO Box 248, Penticton, BC V2A 6J9, Canada\\
$^{12}$National Radio Astronomy Observatory, P.O. Box 0, Socorro, NM 87801, USA\\
$^{13}$School of Physics and Astronomy, University of Southampton, Highfield, Southampton SO17 1BJ, UK\\
$^{14}$New York University Abu Dhabi, P.O. Box 129188, Abu Dhabi, United Arab Emirates\\
$^{15}$Department of Astronomy, University of Virginia, P.O. Box 400325, Charlottesville, VA 22904, USA\\
}
\date{Accepted XXX. Received YYY; in original form ZZZ}

\pubyear{2017}

\begin{document}
\label{firstpage}
\pagerange{\pageref{firstpage}--\pageref{lastpage}}
\maketitle

\begin{abstract}
We present simultaneous radio through sub-mm observations of the black hole X-ray binary (BHXB) V404 Cygni during the most active phase of its June 2015 outburst.  Our $4$ hour long set of overlapping observations with the Very Large Array, the Sub-millimeter Array, and the James Clerk Maxwell Telescope (SCUBA-2), covers 8 different frequency bands (including the first detection of a BHXB jet at $666 \,{\rm GHz}/450\mu m$), providing an unprecedented multi-frequency view of the extraordinary flaring activity seen during this period of the outburst. In particular, we detect multiple rapidly evolving flares, which reach Jy-level fluxes across all of our frequency bands. With this rich data set we performed detailed MCMC modeling of the repeated flaring events. Our custom model adapts the van der Laan synchrotron bubble model to include twin bi-polar ejections, propagating away from the black hole at bulk relativistic velocities, along a jet axis that is inclined to the line of sight. The emission predicted by our model accounts for projection effects, relativistic beaming, and the geometric time delay between the approaching and receding ejecta in each ejection event. We find that a total of 8 bi-polar, discrete jet ejection events can reproduce the emission that we observe in all of our frequency bands remarkably well.  With our best fit model, we provide detailed probes of jet speed, structure, energetics, and geometry. Our analysis demonstrates the paramount importance of the mm/sub-mm bands, which offer a unique, more detailed view of the jet than can be provided by radio frequencies alone.
\end{abstract}



\section{Introduction}
Black hole X-ray binaries (BHXBs), the rapidly evolving, stellar-mass counterparts of active galactic nuclei, are ideal candidates with which to study accretion and accretion-fed outflows, such as relativistic jets. These transient binary systems, containing a black hole accreting mass from a companion star, occasionally enter into bright outburst phases lasting days to weeks, providing a real time view of the evolving relativistic jets (probed by radio through IR frequencies) and accretion flow (probed at X-ray frequencies).

BHXBs display two different types of relativistic jets, dependent on the mass accretion rate in the system \citep{fenbelgal04}. At lower mass accretion rates ($ < 10^{-1} L_{\rm Edd}$)\footnote{The Eddington luminosity is the theoretical limit where, assuming ionized hydrogen in a spherical geometry, radiation pressure balances gravity. This limit corresponds to $L_{\rm Edd} = 1.26\times10^{38}M/M_{\odot} \,{\rm ergs}^{-1}$, where $M$ is the black hole mass.}, during the hard accretion state (see \citealt{remmc06} and \citealt{bel10} for a review of accretion states in BHXBs),
a steady, compact synchrotron-emitting jet is believed to be present in all BHXBs. It has also been shown that the compact jet is not only present during outburst phases, but can persist down into quiescence, at $< 10^{-5} L_{\rm Edd}$ \citep{galc05,plotkin2013,plotkin2015,plotgal15}.  At higher mass accretion rates, during the transition between accretion states, discrete jet ejecta are launched (e.g., \citealt{mirbel4,hjr95,cor02,mj12}), and the compact jet may become quenched \citep{fend99a,cor01,russ11b,corr11h,rush16}. A small number of BHXBs have been observed to display multiple jet ejection events within a single outburst (e.g., \citealt{mirbel4,hjr95,ting95,fen99,kul99,brock2,brock13}).

Compact jets are characterized by a flat to slightly inverted optically thick spectrum ($\alpha>0$; where $f_\nu\propto \nu^\alpha$; \citealt{fen01}), extending from radio up to sub-mm or even infrared frequencies \citep{cor02aa,cas10,tetarenkoa2015}.  
Around infrared frequencies the jet emission becomes optically thin ($\alpha\sim-0.7$; \citealt{rus12}), resulting in a spectral break. Each frequency below this break probes emission (from the optical depth, $\tau=1$ surface) coming from a narrow range of distances downstream in the jet, where higher frequencies originate from regions along the jet axis that are closer to where the jet is launched \citep{blandford79,falcke95}. The exact spectral shape (i.e., spectral index, location of the spectral break) is believed to evolve with changing jet properties such as geometry, magnetic field structure, and particle density profiles \citep{heisun03,mar05,cas09,rus13,van13,rus14}, as well as the plasma conditions in the region where the jet is first accelerated \citep{pol10,pol13,pol14,kolj15}. 

In contrast to the compact jets, jet ejecta are characterized by an optically thin spectrum ($\alpha<0$), give rise to bright flaring activity, and can be routinely resolved with Very Long Baseline Interferometry (VLBI; e.g., \citealt{fend06}). The accompanying flares typically have well defined rise and decay phases, where the flares are usually optically thick in the rise phase, until the self-absorption turnover in the spectrum has passed through the observing band.
These jet ejection events are believed to be the result of the injection of energy and particles to create an adiabatically expanding synchrotron emitting plasma, threaded by a magnetic field (i.e., van der Laan synchrotron bubble model, hereafter referred to as the vdL model; \citealt{vdl66,hj88aa,hjhan95}). In this model, as the source expands
the evolving optical depth results in the distinct observational signature of the lower frequency emission being a smoothed, delayed version of the higher frequency emission.
The ejection events have been linked to both X-ray spectral and timing signatures (e.g., \citealt{fender9,mj12,rus14,kal16}), although a definitive mechanism or sequence of events leading to jet ejection has not yet been identified.
 
Additionally, an extremely rare jet phenomenon, so called jet oscillation events, has also been observed in two BHXBs, GRS 1915$+$105 (radio, mm, IR; \citealt{poolf97}) and V4641 Sgr (optical band; \citealt{uem04}). Such rare events seem to occur only when the accretion rate is at very high fractions of the Eddington rate.
These quasi-periodic oscillations (see \citealt{fenbel04} for a review) show lower frequency emission peaking at later times (consistent with the vdL model for expanding discrete jet ejecta), rise and decay times of the repeated flares that are similar at all frequencies, and time lags between frequencies that vary within a factor of two.  Moreover, no discrete moving components were resolved with VLBI during these oscillation events (although we note this could very well be due to sensitivity limits or the difficulty of synthesis imaging of fast-moving, time-variable components).  As such, the exact nature of these events remains unclear, with theories including discrete plasma ejections, internal shocks in a steady flow, or variations in the jet power in a self-absorbed, conical outflow (e.g., \citealt{fend98,fenpoo00a,col03}). In GRS 1915$+$105 , these oscillations have also been clearly associated with dips in hard X-ray emission, possibly linking the launching of jet ejecta to the ejection and refilling of the inner accretion disc or coronal flow \citep{mir98,bello97,vad01}.

While several transient BHXBs may undergo an outburst period in a given year, in which the jet emission becomes bright enough for detailed multi-wavelength studies, only rare (e.g., once per decade) outbursts probe the process of accretion and the physics of accretion-fed outflows near (or above) the Eddington limit. Observing the brightest and most extreme phases of accretion during these outbursts presents us with a unique opportunity to study jet and accretion physics in unprecedented detail.
On 2015 June 15, the BHXB V404 Cygni entered into one of these rare near-Eddington outbursts. In this paper we report on our simultaneous radio through sub-mm observations of V404 Cygni during the most active phase of this outburst.

 \subsection{V404 Cygni}
V404 Cygni (aka GS 2023$+$338; hereafter referred to as V404 Cyg) is a well studied BHXB that has been in a low-luminosity quiescent state since its discovery with the Ginga satellite in 1989 \citep{ma89}. This source has been observed to undergo a total of three outbursts prior to 2015; most recently in 1989 \citep{hanhj92,ter94,oos97}, and two prior to 1989 which were recorded on photographic plates \citep{r87}. V404 Cyg is known to display bright X-ray luminosities and high levels of multi-wavelength variability, both in outburst and quiescence 
\citep{tanlew,hyn02,hjh89,k89}. The prolonged quiescent period of V404 Cyg, and high quiescent luminosity ($L_X\sim1\times10^{33}\,{\rm erg\,s}^{-1}$; \citealt{corb08}),  has allowed the complete characterization of the system. The optical extinction is low, with $E(B-V) = 1.3$, enabling the study of the optical counterpart, and the determination of the mass function as $6.08\pm0.06\,M_{\odot}$ \citep{casc92,casc94}. Subsequent modelling determined the black hole mass to be $9.0^{+0.2}_{-0.6}\,M_{\odot}$, with an inclination angle of ${{67^\circ}^{+3}_{-1}}$, and an orbital period of 6.5 days \citep{khat10,shab94a}. 
However, we note that this inclination angle estimate is dependent on the assumed level of accretion disc contamination in the optical light curves being modelled. \cite{khat10} assumed $<3\%$ accretion disc contamination, but given that V404 Cyg is known to be variable in quiescence in the optical, it is plausible that the accretion disc contamination may be larger \citep{zur03,bern16}, which would imply a larger inclination angle.
Further, the faint, unresolved radio emission from the quiescent jets was used to determine a model-independent parallax distance of $2.39\pm0.14$ kpc \citep{mj9}, making V404 Cyg one of the closest known BHXBs in the Galaxy. The close proximity,
well-determined system parameters, and bright multi-wavelength activity make this system an ideal target for jet and accretion studies.

On 2015 June 15\footnote{\cite{bern16} serendipitously detected an optical precursor to this outburst on June 8/9, approximately one week prior to the first X-ray detection.}, V404 Cyg entered into its fourth recorded outburst period. The source began exhibiting bright multi-wavelength 
flaring activity (e.g., \citealt{ferrignoc15,gandhip15,gazeask15,mooleyk15,mottas15a,mottas15b,tetarenkoa15,tetarenkoa15b}) immediately following the initial detection of the outburst in X-rays \citep{barth15,negoro2015b, kuulkerse15}, and swiftly became the brightest BHXB outburst seen in the past decade. This flaring behaviour was strikingly similar to that seen in the previous 1989 outburst \citep{ter94,oos97,z99}. Towards the end of June the flaring activity began to diminish across all wavelengths (e.g., \citealt{ferrignoc15b,martincarilloa15b,oatessr15,scarpacij15b,tetarenkoa15c,tsubonok15b}), and the source began to decay \citep{sivakoffgr15b,sivakoffgr15d}, reaching X-ray quiescence\footnote{V404 Cyg entered optical quiescence in mid October 2015 \citep{bern16a}.}  in early to mid August \citep{sivakoffgr15e,plot16}. V404 Cyg also showed a short period of renewed activity from late December 2015 to early January 2016 (e.g., \citealt{lip15,trus15,beard15,mal15,tetarenkoa16a,mott16a}), and \cite{mun16b} present radio, optical, and X-ray monitoring during this period.

We organized simultaneous observations with the Karl G. Jansky Very Large Array (VLA), the Sub-millimeter Array (SMA), and the James Clerk Maxwell Telescope (JCMT) on 2015 June 22 (approximately one week following the initial detection of the outburst), during which time some of the brightest flaring activity seen in the entire outburst was observed. 
This comprehensive data set gives us an unprecedented multi-frequency view of V404 Cyg, in turn allowing us to perform detailed multi-frequency light curve modelling of the flaring events. In \S 2 we describe the data collection and data reduction processes. \S 3 describes the custom procedures our team developed to extract high time resolution measurements from our data. In \S 4 we present our multi-frequency light curves, outline our model, and describe the modelling process.  A discussion of our best fit model is presented in \S 5, and a summary of our work is presented in \S 6.

\section{Observations and Data Analysis}
\subsection{VLA Radio Observations}
We observed V404 Cyg with the VLA (Project Code: 15A-504) on 2015 June 22, with scans on source from 10:37:24--14:38:39 UTC (${\rm MJD}=57195.442-57195.610$) in both C ($4-8\,{\rm GHz}$) and K ($18-26\,{\rm GHz}$) band. The array was in its most extended A configuration,
where we split the array into 2 sub-arrays of 14 (sub-array A) and 13 (sub-array B) antennas. Sub-array A observed the sequence C-K-C, while sub-array B observed the sequence K-C-K, with an 80 second on target and 40 second on calibrator cycle, in order to obtain truly simultaneous observations across both bands. All observations were made with an 8-bit sampler, comprised of 2 base-bands, with 8 spectral windows of 64 2 MHz channels each, giving a total bandwidth of 1.024 GHz per base-band. Flagging, calibration, and imaging of the data were carried out within the Common Astronomy Software Application package (CASA; \citealt{mc07}) using standard procedures. We used 3C48 (0137$+$331) as a flux calibrator, and J2025$+$3343 as a phase calibrator for both sub-arrays. No self-calibration was performed. 
Due to the rapidly changing flux density of the source, we imaged the source (with natural weighting; {see the Appendix for details on our choice of weighting scheme}) on timescales as short as the correlator dump time (2 seconds) using our custom CASA timing scripts (see \S 3.1 for details).

\begin{figure*}
\begin{center}
 \includegraphics[width=1.8\columnwidth]{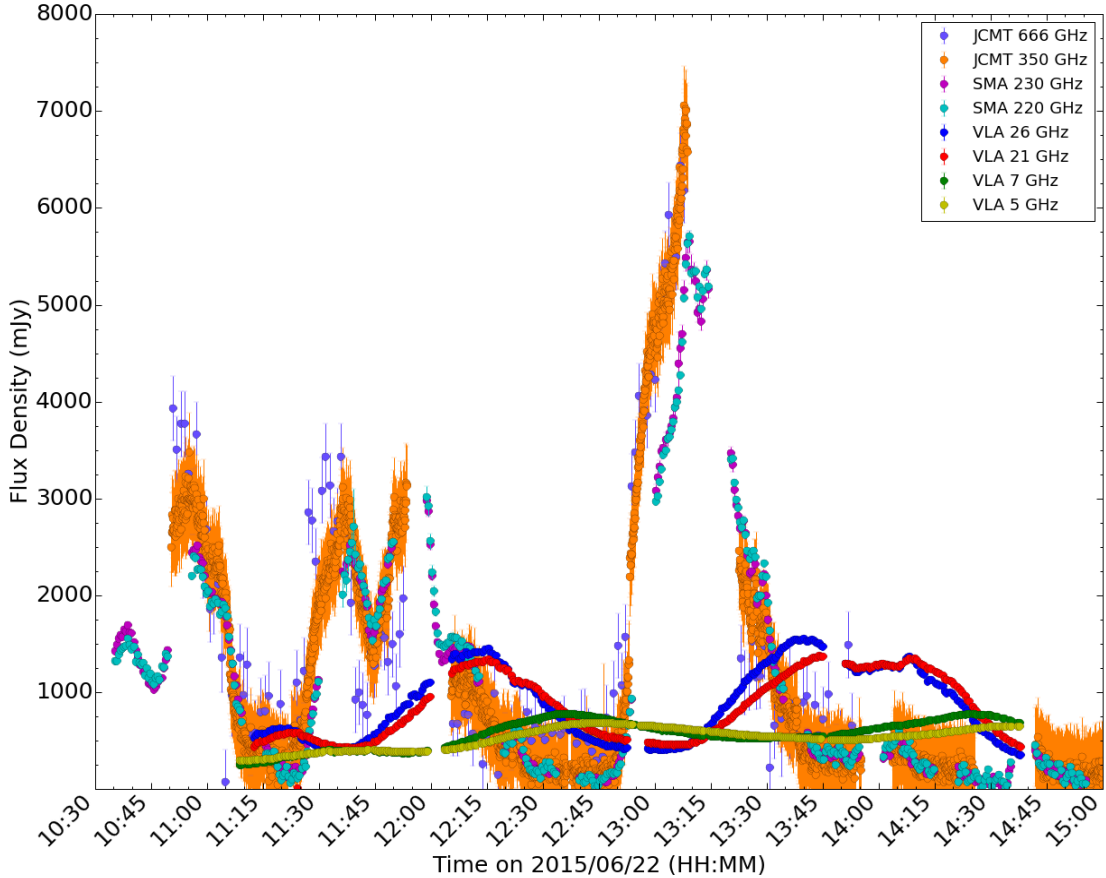}
 \caption{\label{fig:lc}  \small Simultaneous radio through sub-mm light curves of the BHXB V404 Cygni during the most active phase of its June 2015 outburst. 
 These light curves sample the brightest flares at these frequencies over the entire outburst. All light curves are sampled at the finest time resolution possible, limited only by the correlator dump time (and the sensitivity for JCMT data). The VLA light curves have 2 second time bins, the SMA light curves have 30 second time bins, the JCMT SCUBA-2 350 GHz ($850\,\mu m$) light curve has 5 second time bins, and the JCMT SCUBA-2 666 GHz ($450\,\mu m$) light curve has 60 second time bins. The mm/sub-mm regime samples a much more extreme view of the flaring activity than the radio regime, with detailed sub-structure detected only in the mm/sub-mm light curves.
}
\end{center}
\end{figure*}

\subsection{SMA (Sub)-Millimetre Observations}
We observed V404 Cyg with the SMA (Project Code: 2015A-S026) on 2015 June 22, with scans on source from 10:16:17--18:20:47 UTC (${\rm MJD}=57195.428-57195.764$), and the correlator tuned to an LO frequency of 224 GHz. The array was in the sub-compact configuration with a total of 7 antennas (out of a possible 8 antennas). These observations were made with both the ASIC and SWARM \citep{swarm} correlators active, to yield 2 side-bands, with 48 spectral windows of 128 0.8125 MHz channels (ASIC) and an additional $2$ $1.664$ GHz spectral windows (SWARM), giving a total bandwidth of 8.32 GHz per side-band. The SWARM correlator had a fixed resolution of 101.6 kHz per channel, and thus originally 16383 channels for each SWARM spectral window.  Given the continuum nature of these observations, we performed spectral averaging, to yield 128 13 MHz channels in both SWARM spectral windows, to match the number of channels in the ASIC spectral windows, and in turn make it easier to combine ASIC and SWARM data. 
We used 3C454.3 (J2253$+$1608) as a bandpass calibrator, MWC349a and J2015$+$3710 as phase calibrators, and Neptune and Titan as flux calibrators\footnote{The SMA calibrator list can be found at http://sma1.sma.hawaii.edu/callist/callist.html.}. We note that only the second IF (spectral windows 25-50) was used for flux calibration in the upper side-band due to a CO line that was present in both flux calibrators at 230.55 GHz.
Our observing sequence consisted of a cycle of 15 min on target and 2.5 min on each of the two phase calibrators. As CASA is unable to handle SMA data in its original format, prior to any data reduction we used the SMA scripts, sma2casa.py and smaImportFix.py, to convert the data into CASA MS format, perform the ${\rm T}_{\rm sys}$ corrections, and spectrally average the two SWARM spectral windows.
Flagging, calibration, and imaging of the data were then performed in CASA using procedures outlined in the CASA Guides for SMA data reduction\footnote{Links to the SMA CASA Guides and these scripts are publicly available at https://www.cfa.harvard.edu/sma/casa.}. Due to the rapidly changing flux density of the source, we imaged the source (with natural weighting; {see the Appendix for details on our choice of weighting scheme}) on timescales as short as the correlator dump time (30 seconds) using our custom CASA timing scripts (see \S 3.1 for details).

\subsection{JCMT SCUBA-2 (Sub)-Millimetre Observations}
We observed V404 Cyg with the JCMT (Project Code: M15AI54) on 2015 June 22 from 10:49:33--15:12:40 UTC (${\rm MJD}=57195.451-57195.634$), in the $850\mu m$ (350 GHz) and $450\mu m$ (666 GHz) bands. The observation consisted of eight $\sim30$ min scans on target with the SCUBA-2 detector \citep{chap,holl}. To perform absolute flux calibration, observations of the calibrator CRL2688 were used to derive a flux conversion factor \citep{demp}. The daisy configuration was used to produce 3 arcmin maps of the target source region. During the observations we were in the Grade 3 weather band with a 225 GHz opacity of 0.095--0.11. Data were reduced in the StarLink package using both standard procedures outlined in the SCUBA-2 cookbook\footnote{http://starlink.eao.hawaii.edu/devdocs/sc21.htx/sc21.html} and SCUBA-2 Quickguide\footnote{https://www.eaobservatory.org/jcmt/instrumentation/continuum\\
/scuba-2/data-reduction/reducing-scuba2-data}, as well as a custom procedure to create short timescale maps (timescales shorter than the 30 minute scan timescale) to extract high time resolution flux density measurements of the rapidly evolving source (see \S 3.2 for details).

\section{High Time Resolution Measurements}
\subsection{VLA and SMA}
To obtain high time resolution flux density measurements of V404 Cyg from our interferometric data sets (VLA and SMA) we developed a series of custom scripts that run within CASA.  A detailed account of the development and use of these scripts will be presented in Tetarenko \& Koch et al. 2017, in prep., although we provide a brief overview of the capabilities here.

Our scripts split an input calibrated CASA Measurement Set into specified time intervals for analysis in the image plane or the uv plane. In the image plane analysis, each time interval is cleaned and the flux density of the target source is measured by fitting a point source in the image plane with the native CASA task \texttt{imfit}. All imaging parameters (e.g., image size, pixel size, number of CLEAN iterations, CLEAN threshold) can be fully specified. In the uv plane analysis, the \textsc{uvmultfit} package \citep{mart14} is used to measure flux density of the target source. In either case, an output data file and plot of the resulting light curve are produced.
These scripts are publicly available on github\footnote{https://github.com/Astroua/AstroCompute\_Scripts}, and are being implemented as a part of an interactive service our team is developing to run on Amazon Web Services Cloud Resources.

All VLA and SMA flux density measurements output from this procedure (fitting only in the image plane) are provided in a machine readable table online, which accompanies this paper. Additionally, to check that the variability we observed in V404 Cyg is dominated by intrinsic variations in the source and not due to atmospheric or instrumental effects, we also ran our calibrator sources through these scripts (see the Appendix for details).

\subsection{JCMT SCUBA-2}
To obtain high time resolution flux density measurements of V404 Cyg from our JCMT SCUBA-2 data we developed a custom procedure to produce a data cube, containing multiple maps of the target source region, at different time intervals throughout our observation.

We run the StarLink Dynamic Iterative Mapmaker tool on each of the target scans, using the bright compact recipe, with the addition of the {shortmap} parameter. The {shortmap} parameter allows the Mapmaker to create a series of maps, each of which will include data from a group of adjacent time slices. The number of time slices included in each map is equivalent to the shortmap parameter value. At $850\mu m$ we use ${\rm shortmap}=200$ to produce 362 time slices for a 32 minute scan, resulting in 5 second time bins. At $450\mu m$ ${\rm shortmap}=400$ would produce the same number of time slices, where a factor of 2 is applied as the default pixel size is 2 arcsec at $450 \mu m$ and 4 arcsec at $850\mu m$. However, as the noise is higher at $450 \mu m$, we use ${\rm shortmap}=4800$ to produce 32 time slices for a 32 minute scan, resulting in 60 second time bins.
The \texttt{stackframes} task is then used to combine all of the short maps into a cube for each scan. The ${\rm sort}={\rm True}$ and ${\rm sortby}={\rm MJD-AVG}$ parameters ensure the maps are ordered chronologically in time, with the resulting cube having the dimensions, position X (pixels), position Y (pixels), time (MJD).
Using the \texttt{wcsmosaic} task we then combined the cubes from all the scans.
We calibrated the combined cube into units of Jy using the \texttt{scuba2checkcal} and \texttt{cmult} tasks. Finally, the combined cube can be viewed in \texttt{Gaia}, and converted to FITS format with the \texttt{ndf2fits} task.

\begin{figure}
\begin{center}
 \includegraphics[width=1\columnwidth]{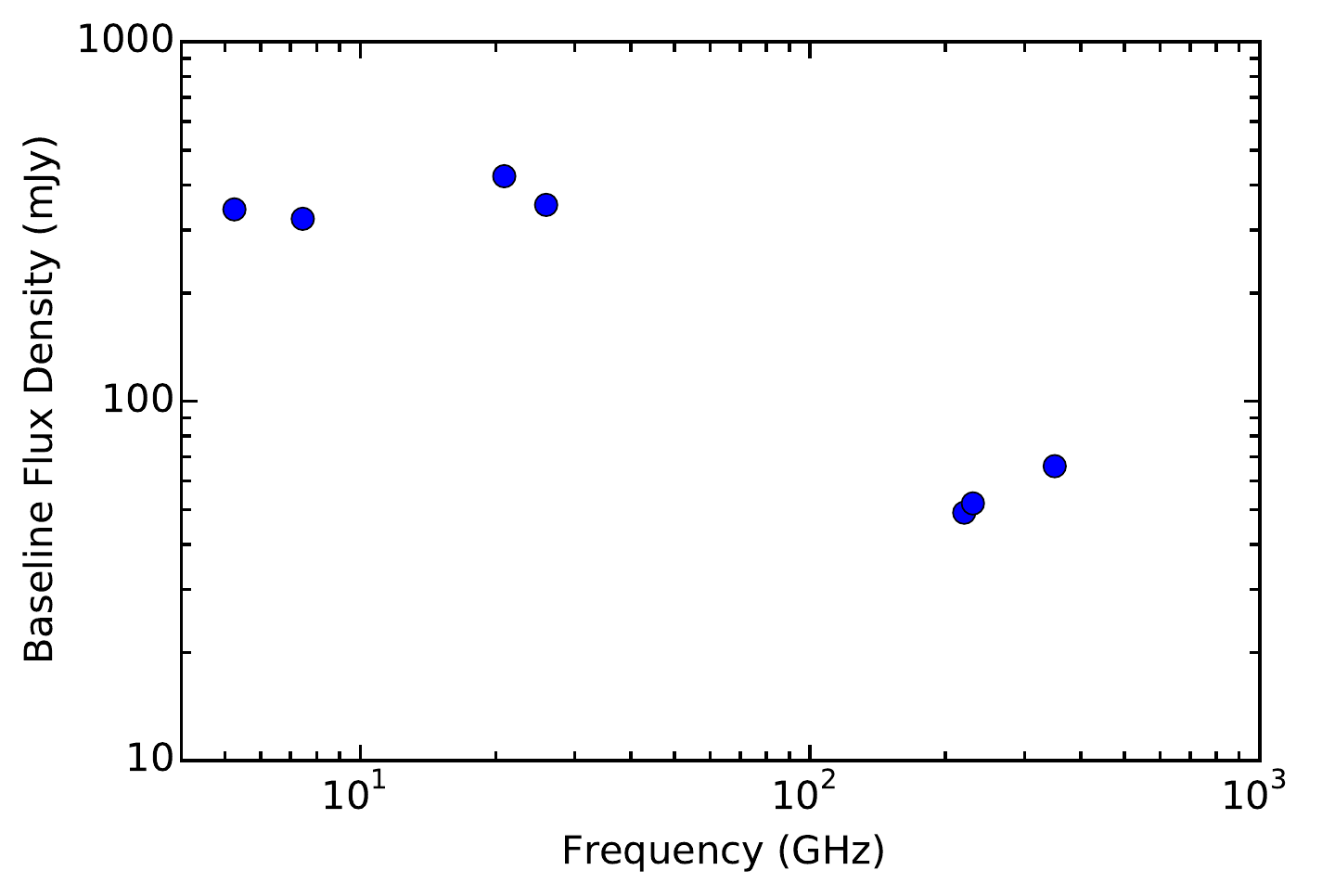}
\caption{\label{fig:blf}  \small Estimated radio through sub-mm spectrum of the baseline flux component seen in our light curves.
}
\end{center}
\end{figure}

To extract flux densities from each time slice in the combined cube, we fit a 2D gaussian\footnote{The python package \textsc{gaussfitter} is used in the gaussian fitting; https://github.com/keflavich/gaussfitter} with the size of the beam (FWHM of 15.35 arcsec at $850\mu m$ and 10.21 arcsec at $450 \mu m$; derived using the task \texttt{scuba2checkcal}) to each slice of the cube.
All JCMT SCUBA-2 flux density measurements output from this procedure are provided in a machine readable table online, which accompanies this paper. As with our interferometric data sets, to check that the variability we observed in V404 Cyg is dominated by intrinsic variations in the source and not due to atmospheric or instrumental effects, we also ran this procedure on our calibrator source scans (see the Appendix for details).

\section{Results}
\subsection{Multi-frequency Light Curves}
A composite light curve of all of our VLA, SMA and JCMT observations from June 22 is presented in Figure~\ref{fig:lc}. We observe rapid multi-frequency variability in the form of multiple large scale flares, reaching Jy flux levels. 
 In the SMA data, the largest flare (at $\sim$ 13:15 UTC) rose from $\sim 100$ mJy  to a peak of $\sim 5.6$ Jy on a timescale of $\sim 25$ min. The JCMT SCUBA-2 data appear to track the SMA data closely, with the largest flare at $350$ GHz rising from $\sim400$ mJy to a peak of $\sim 7.2$ Jy on a timescale of $\sim18$ min. This is the largest mm/sub-mm flare ever observed from a BHXB, far surpassing even the brightest events in GRS 1915$+$105 \citep{fenpoo00a}. The VLA radio data lag the mm/sub-mm (where the lag appears to be variable among the flares; $\sim$20--45 min \& $\sim$40--75 min between 350 GHz and the 18--26 GHz \& 4--8 GHz bands, respectively), with flares in the 18--26 GHz band rising to a peak of $\sim1.5$ Jy on a timescale of $\sim35$ min, and flares in the 4--8 GHz band rising to a peak of $\sim 780$ mJy on a timescale of $\sim45$ min.

\begin{figure}
\begin{center}
 \includegraphics[width=0.8\columnwidth]{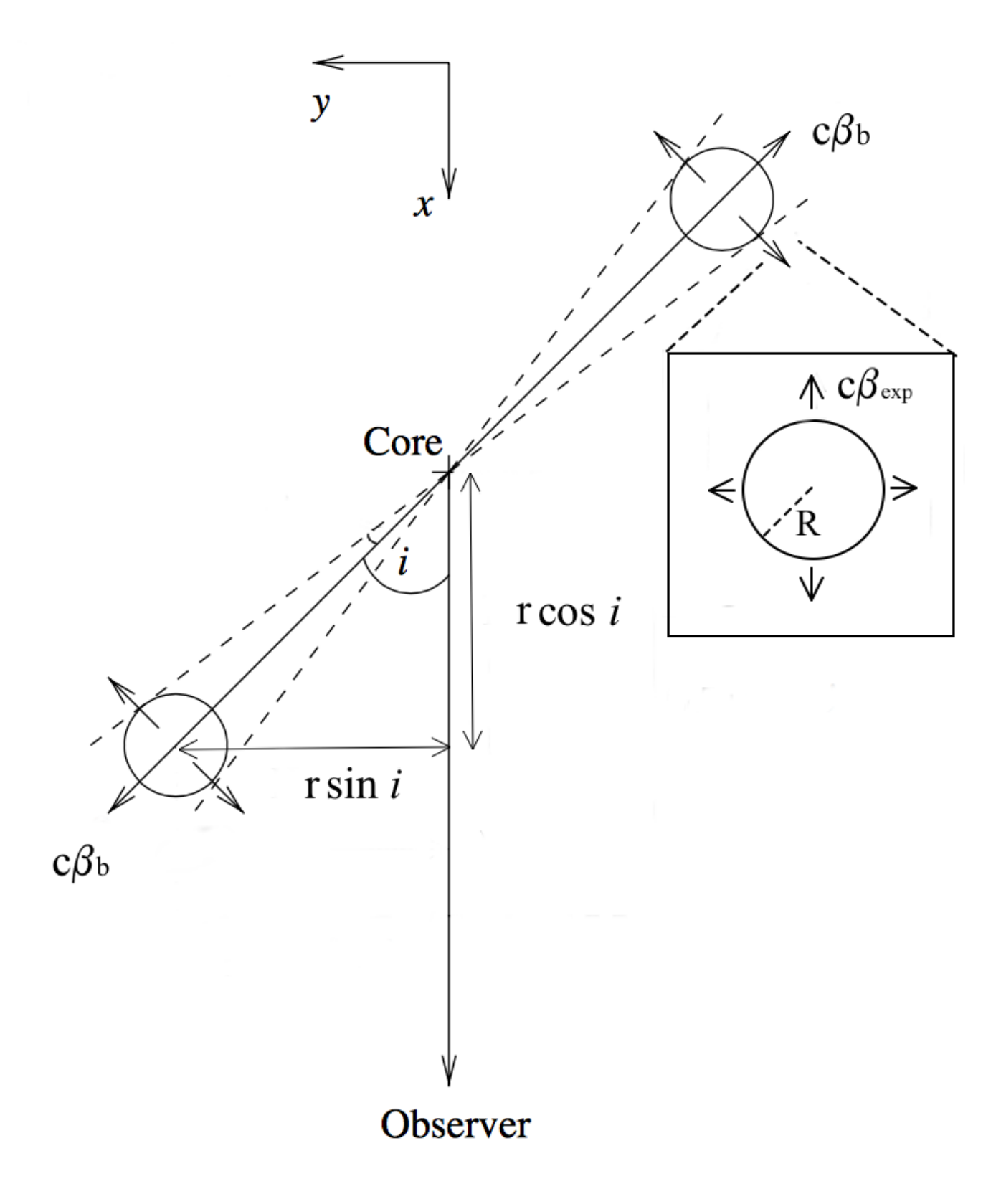}
 \caption{ \small Schematic of the geometry for our jet model in a plane defined by our line-of-sight and the central axis of the jets
  (i.e., bird's eye view).  The inset panel displays the ejecta component seen from the source frame (at rest with respect to the ejecta). All parameters are defined within the accompanying text. This figure was adapted from its original form in \citealt{millerj06}.}
\label{fig:mj1}
\end{center}
\end{figure}

Upon comparing the multi-frequency emission, it is clear that the mm/sub-mm data provide a much more extreme view of the flaring activity than the radio emission. In particular, there is more structure present in the mm/sub-mm light curves when compared to the radio light curves. As such, while not immediately apparent in the radio light curves, the mm/sub-mm data suggest that each of the three main flares in the light curves is actually the result of the superposition of emission from multiple flaring components. 
Additionally, the lower frequency emission in the light curves appears to be a smoothed, delayed version of the high frequency emission (with the flares showing longer rise times at lower frequencies).
This emission pattern is consistent with an expanding outflow structure, where the mm/sub-mm emission originates in a region (with a smaller cross-section) closer to the black hole, and has thus not been smoothed out to as high a degree as the radio emission, as the material expands and propagates outwards. 
Therefore, all of these observations suggest that the emission in our light curves could be dominated by emission from multiple, expanding, discrete jet ejection events \citep{vdl66}.

Further, we notice that the baseline flux level at which the flaring begins at each frequency in our light curves appears to vary. 
This suggests that there is an additional frequency-dependent component contributing to our light curves, on top of the discrete jet ejecta.
In an effort to determine the origin of this extra emission, assuming that the baseline emission is constant in time, we create a spectrum of this emission by estimating the baseline flux level at each frequency (we performed iterative sigma clipping and take the minimum of the resulting sigma clipped data). This spectrum\footnote{We note that these are only empirical initial estimates of the baseline flux at each frequency, and do not necessarily represent the flux of the compact jet in our model presented in \S 4.2.} is presented in Figure~\ref{fig:blf}, {where it appears as though the baseline emission could be described by a broken power-law or a single power-law (with higher frequency emission displaying a lower baseline level than lower frequency emission).} This spectral shape, combined with the fact that we observe a strong compact core component (in addition to resolved ejecta components) within simultaneous high resolution radio imaging (Miller-Jones, et al. 2017, in prep), suggests that the baseline emission originates from an underlying compact jet that was not fully quenched.

 \begin{figure}
\begin{center}
 \includegraphics[width=0.7\columnwidth]{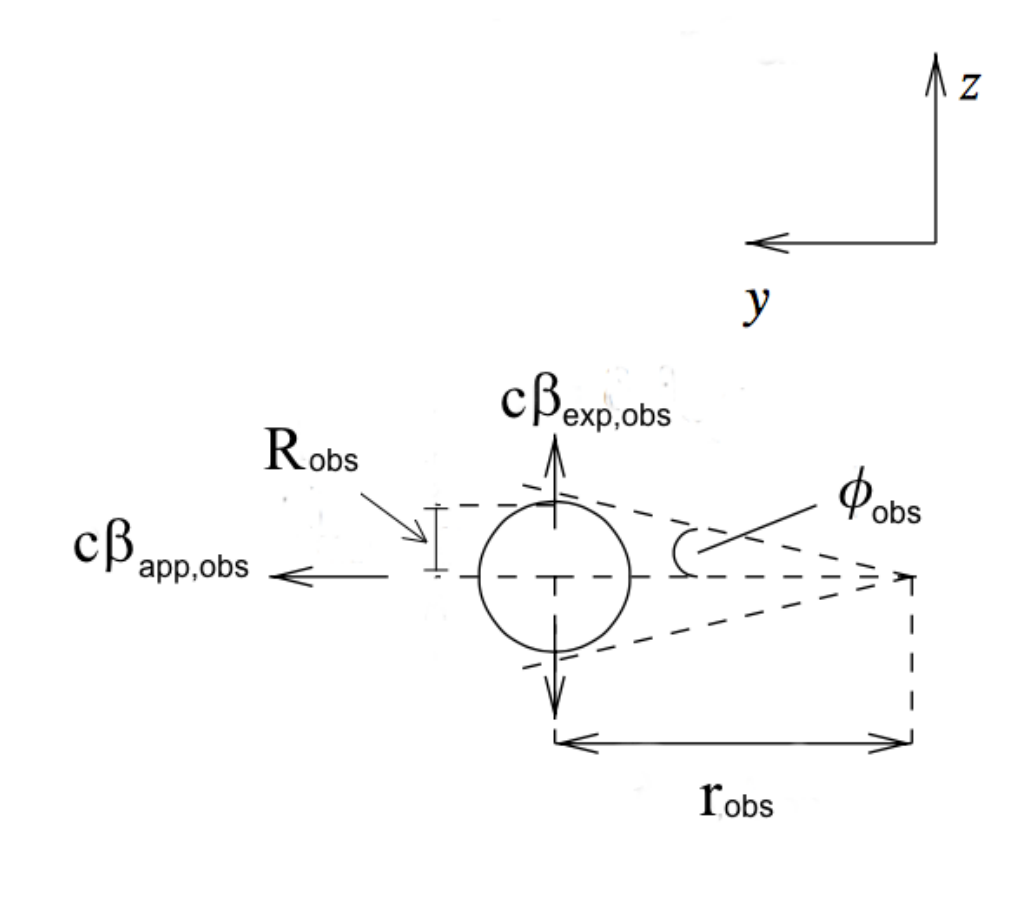}
 \caption{\label{fig:mj2}  \small Schematic of the geometry of the discrete jet ejections in our model, as seen by the observer. All parameters are defined within the accompanying text. This figure was adapted from its original form in \citealt{millerj06}.
}
\end{center}
\end{figure}

\subsection{V404 Cyg Jet Model}
Given the morphology of our light curves outlined in the previous section, we have constructed a jet model for V404 Cyg that is capable of reproducing emission from multiple, repeated, discrete jet ejection events, on top of an underlying compact jet component. We define two coordinate frames, the observer frame and the source frame (at rest with respect to the ejecta components). We will compute our model primarily in the source frames, and then transform back to the observer frame. All variables with the subscript \textit{obs} are defined in the observer frame.
Schematics displaying the geometry of our model from different viewpoints are displayed in Figures~\ref{fig:mj1} \& \ref{fig:mj2}.
 
{In our model, the underlying compact jet is characterized by a broken power-law spectrum, where the flux density is independent of time and varies only with frequency according to,
\begin{equation}
F_{\nu,{\rm cj}} = \left \{
                     \begin{array}{ll}
                       F_{{\rm br},{\rm cj}} (\nu / \nu_{\rm br}) ^ {\alpha_1} & ,\, \nu < \nu_{\rm br} \\
                       F_{{\rm br},{\rm cj}} (\nu / \nu_{\rm br}) ^ {\alpha_2} & , \, \nu > \nu_{\rm br} \\
                     \end{array}
                   \right.
\end{equation}
Here $\nu_{\rm br}$ represents the frequency of the spectral break, $F_{{\rm br,cj}}$ represents the amplitude of the compact jet at the spectral break frequency, $\alpha_1$ represents the spectral index at frequencies below the break, and $\alpha_2$ represents the spectral index at frequencies above the break. In the case where the spectral break frequency is located below the lowest sampled frequency band, or above the highest sampled frequency band, the underlying compact jet can be characterized by a single power-law spectrum, where $F_{\nu,{\rm cj}}=F_{0,{\rm cj}}\left(\frac{\nu}{\nu_0}\right)^{\alpha}$. Here $F_{{\rm 0,cj}}$ represents the amplitude of the compact jet at $\nu_{\rm 0}$, and $\alpha$ represents the spectral index.}

On top of the compact jet, we define a discrete ejection event as the simultaneous launching of two identical, bi-polar plasma clouds (an approaching and receding component). 
Each of these clouds evolve according to the vdL model \citep{vdl66}. In this model, a population of relativistic electrons, with a power-law energy distribution ($N(E)dE=KE^{-p}dE$), is injected into a spherical cloud threaded by a magnetic field.
The cloud is then allowed to expand adiabatically, while the electrons and magnetic field are assumed to be kept in equipartition. 
As a result of the expansion, this model predicts the flux density of each cloud will scale as,
\begin{equation}
F_{\nu,{\rm ej}}=F_0 \left(\frac{\nu}{\nu_0}\right)^{5/2} \left(\frac{R}{R_0}\right)^{3}\frac{1-\exp(-\tau_\nu)}{1-\exp(-\tau_0)}.
\end{equation}
Here $R$ indicates the time-dependent radius of the cloud, and the synchrotron optical depth, $\tau_\nu$, at a frequency, $\nu$, scales as,
\begin{equation}
\tau_\nu=\tau_0 \left(\frac{\nu}{\nu_0}\right)^{-(p+4)/2} \left(\frac{R}{R_0}\right)^{-(2p+3)}.
\end{equation}
Note that the subscript 0 in all our equations indicates values at the reference frequency\footnote{We defined our reference frequency as the upper-sideband in our SMA data (230 GHz).}, at the time (or radius) of the peak flux of the component.

Taking the derivative of Equation 2 with respect to time\footnote{Our expression in Equation 4 differs from that of \cite{vdl66}, as he takes the derivative with respect to $\nu$ instead of time, yielding ${\rm e}^{\tau_0}-([p+4]/5)\tau_0-1=0$.} (or radius), allows us to relate the optical depth at which the flux density of the reference frequency reaches a maximum, $\tau_0$, to the power-law index of the electron energy distribution, $p$,
\begin{equation}
e^{\tau_0}-(2p/3+1)\tau_0-1=0.
\end{equation}

Equation 4 has no analytic solution and thus must be solved numerically. Therefore, we choose to leave our model in terms of $\tau_0$, and solve for $p$ after the fitting process.

To describe the time-dependence of the cloud radius, a linear expansion model is used, according to,
\begin{equation}
R=R_0+\beta_{\rm exp}c\left(t-t_0\right).
\end{equation}
Here $\beta_{\rm exp}c$ represents the expansion velocity of the cloud, while $R_0$ can be expressed in terms of the distance to the source, $d$, peak flux, $F_0$, and optical depth, $\tau_0$, of the cloud at the reference frequency \citep{yus08},
\begin{equation}
R_0=\left[\frac{F_0d^2}{\pi}\frac{1}{1-{\rm exp}(-\tau_0)}\right]^{1/2}.
\end{equation}

At the same time that the clouds are expanding, they are also propagating away from the black hole at bulk relativistic velocities, along a jet axis that is inclined to the observer's line of sight (see Figure~\ref{fig:mj1}). As such, the emission we observe will have been affected by projection effects, relativistic beaming, and a geometric time delay between the approaching and receding clouds in each ejection event. 

To account for these effects, we first assume that the clouds are travelling at a constant bulk velocity, $\beta_b c$, and that the jet has a conical geometry (with an observed opening angle, $\phi_{\rm obs}$). In turn, the apparent observed velocity across the sky (derived via the transverse Doppler effect) is represented as \citep{mirabelrodriguez99},

\begin{equation}
\beta_{\rm app,obs} = \left \{
                     \begin{array}{ll}
                       \frac{r{\rm sin}\,i}{c\,(t-t_{\rm ej}) - r{\rm cos}\,i}& \rightarrow\, {\rm approaching}\\
                       &\hspace{2.5cm},\\
                       \frac{r{\rm sin}\,i}{c\,(t-t_{\rm ej}) + r{\rm cos}\,i} & \rightarrow \, {\rm receding}\\
                     \end{array}
                   \right.
\end{equation}

\noindent where $r=\beta_b c (t-t_{\rm ej})$ is the distance travelled by the cloud away from the black hole, $t_{\rm ej}$ represents the ejection time, $c$ represents the speed of light, and $i$ represents the inclination angle of the jet axis to the line of sight.

Equation 7 can be simplified by substituting in our expression for $r$ to yield,
\begin{equation}
\beta_{\rm app,obs}=\beta_b\Gamma\delta_\mp{\rm sin}(i),
\end{equation}
where the Doppler factor and bulk Lorentz factor are given by $\delta_\mp=\Gamma^{-1}[1\mp\beta_b{\rm cos}\,i]^{-1}$ and $\Gamma=(1-\beta_b^2)^{-1/2}$, respectively. The sign convention in the Doppler factor indicates that a $\delta_{-}$ should be used for the approaching cloud and a $\delta_{+}$ should be used for the receding cloud.

From Figures~\ref{fig:mj1} \& \ref{fig:mj2},
\begin{equation}
\footnotesize
{\rm tan}\,\phi_{\rm obs}=\frac{R_{\rm obs}}{r_{\rm obs}}=\frac{\delta_\mp\,\beta_{\rm exp}\,c\,(t-t_{ej})_{\rm obs}}{\beta_{\rm app,obs}\,c\,(t-t_{\rm ej})_{\rm obs}}=\frac{\delta_\mp\,\beta_{\rm exp}}{\beta_{\rm app,obs}}.
\end{equation}

Combining Equations 8 \& 9, and solving for the bulk Lorentz factor, $\Gamma$, yields,
\begin{equation}
\Gamma={\left(1+\frac{\beta_{\rm exp}^2}{{\rm tan}^2\phi_{\rm obs}{\rm sin}^2i}\right)}^{1/2}.
\end{equation}

Rearranging Equation 10 (and substituting in $1-\Gamma^2=-\Gamma^2\beta_b^2$) gives the expansion velocity (to be input into Equation 5) in terms of only the bulk velocity and jet geometry (inclination and opening angle), such that,
\begin{equation}
\beta_{\rm exp}={\rm tan}\,\phi_{\rm obs}[{\Gamma^2\{1-{(\beta_b {\rm cos\,}i)}^2\}-1]}^{1/2}.
\end{equation}

\renewcommand\tabcolsep{2.5pt}
 \begin{table*}
\caption{V404 Cyg Jet Model Parameters and Priors$^a$}\quad
\centering
\begin{tabular}{ lllcc }
 \hline\hline
  {\bf Parameter$^b$} &{\bf Description}&{\bf Prior Distribution}$^c$&{\bf Prior}&{\bf Prior}\\
  &&&{\bf Minimum}&{\bf Maximum}\\[0.15cm]
  \hline
$F_{0,\rm cj}$ &amplitude of compact jet component at $\nu_0$ in mJy& uniform&10&1000\\[0.1cm]
$\alpha$ &spectral index& truncated normal ($\mu=-0.5$, $\sigma=0.1$)&-1&0\\[0.1cm]
$^*$$t_{\rm ej}$ &ejection time of bi-polar components in seconds$^d$$^,$$^e$& uniform&$t_g-1000$ & $t_g+1000$\\[0.1cm]
$^*$$i$ &inclination angle of the jet axis in degrees& truncated normal ($\mu=67$, $\sigma=20$)&0&90\\[0.1cm]
$^*$$\phi_{\rm obs}$&observed opening angle in degrees& uniform&1&20\\[0.1cm]
$^*$$\tau_0$ &optical depth at the reference frequency& truncated normal ($\mu=2.0$, $\sigma=0.3$)&1&3\\[0.1cm]
$^*$$F_{0}$ &peak flux density at the reference frequency in mJy& truncated normal ($\mu=3000$, $\sigma=1000$)&0&6000\\[0.1cm]
$^*$$\beta_b$&bulk jet speed in units of c& inverse gamma ($a=3$)$^f$&0&1\\[0.15cm] \hline
\end{tabular}\\
\begin{flushleft}
{$^a$ {Note that the emission from the underlying compact jet portion of our model is best fit by a single power-law rather than a broken power-law. Therefore, only the parameters describing the single power-law version of the compact jet in our model are shown here.}}\\
{$^b$Parameters marked with a $^*$ indicate those which are allowed to vary between ejection events.}\\
{$^c$Justification of our use of these priors is presented in the text of \S4.3.}\\
{$^d$For simplicity, when modelling we convert our times to units of seconds past the zero point of MJD 57195.41806.}\\
{$^e$$t_g$ represents the initial guess of the ejection time, see \S 4.3 for details.}\\
{$^f$The inverse gamma distribution takes the form, $f(x)=\frac{x^{(-a-1)}} { \Gamma(a)}{\rm exp}(-1/x)$, where $\Gamma$ represents the gamma function not the bulk Lorentz factor. This distribution is a common prior used for small number parameters defined to be $<1$.}\\
\end{flushleft}
\label{table:params}
\end{table*}
\renewcommand\tabcolsep{6pt}

Further, we wish to write our model in terms of only the ejection time ($t_{\rm ej}$), rather than the time of the peak flux at the reference frequency ($t_0$), without introducing any additional parameters. Using our definition that $R=R_0$ at the instant $t=t_0$, the two timescales are related by,
\begin{equation}
t_0=t_{\rm ej}+\frac{R_0}{\beta_{\rm exp}c}.
\end{equation}

Lastly, we correct for relativistic beaming by applying a factor of $\delta_\mp^3$ \citep{long11} to our flux density in Equation 2, according to,
\begin{equation}
F_{\nu,{\rm ej,obs}}=\delta_\mp^3F_{\nu,{\rm ej}}.
\end{equation}

Here $F_{\nu,{\rm ej,obs}}$ indicates the flux density of the cloud in the observer frame, at the observing frequency $\nu_{\rm obs}$, at the observed times since the zero point of our observations, $\Delta t_{\rm obs}$, while $F_{\nu,{\rm ej}}$ indicates the flux density of the clouds in the source frame, at the frequency, $\nu=\delta_\mp^{-1}\nu_{\rm obs}$, at the times, $\Delta t=\delta_\mp \Delta t_{\rm obs}$.

All of the ejection events we model are not correlated, 
and thus evolve independently of each other. The total observed flux density in our model is represented as,
\begin{equation}
\small
F_{\nu,{\rm obs,tot}}=\sum_i\delta_{-}^3(F_{\nu,{\rm i},{\rm app}})+\sum_i\delta_{+}^3(F_{\nu,{\rm i},{\rm rec}})+F_{\nu,{\rm cj}}.
\end{equation}

\subsubsection{Jet Precession}
In addition to our VLA, SMA, and JCMT observations, we also obtained simultaneous high angular resolution radio observations with the Very Long Baseline Array (VLBA). Through imaging the VLBA data set in short 2 minute time bins, we resolve multiple discrete ejecta. Our analysis of these VLBA images has shown clear evidence of jet precession, where the position angle of the resolved ejecta change by up to 40 degrees on an hourly timescale (this result will be be reported in detail in Miller-Jones et al. 2017, in prep.). As the emission predicted by our model is highly dependent on the inclination angle of the jet axis, we account for the effect of this rapid, large scale jet precession in our model by allowing our inclination parameter, $i$, to vary between ejection events.

\subsubsection{Accelerated Motion}
While we have assumed that the jet ejecta are travelling at constant bulk velocities, it is possible that they undergo some form of accelerated motion. To test this hypothesis we generalized our model to allow the input of a custom bulk velocity profile, where we implemented simple velocity profiles to mimic a finite acceleration period where the cloud would approach a terminal velocity (e.g., a linear ramp function, a body subject to a quadratic drag force). However, in all cases, our best fit model either tended towards a constant velocity profile, or would not converge. This result, while not ruling out the possibility of accelerated motion, suggests that any potential acceleration period may have only lasted for a short enough period of time that we are not able to discern the difference between the resulting light curves for the accelerated and constant bulk motion. 

\subsubsection{Sub-Conical Jet Geometry}
While we have assumed that the jet in our model is conical (constant opening angle), it is possible that the jet geometry could deviate from a strictly conical shape (especially on the AU size scales we are probing), where the opening angle (and in turn the expansion speed of the ejecta) could change with time. In particular, if we assume that the jet confinement mechanism is external, then the jet geometry will depend on the adiabatic indices of the two media (i.e., the jet and its surrounding medium). A relativistic plasmon confined by the internal pressure of a terminal spherical wind (made up of a $\Gamma=5/3$ gas) will expand sub-conically, according to $R \propto r^{5/6}$. To test this scenario, we modified our model to use the above sub-conical expansion expression in place of Equation 5. In doing this we find that our best fit model still tends toward constant expansion speed/opening angle profiles for all the ejecta. This result, while not ruling out a non-conical jet geometry, could suggest that any deviations from a conical jet shape only occur on sub-AU size scales, probing timescales before the sub-mm emission peaks, and thus we are not able to discern the difference between the resulting light curves for conical/sub-conical jet geometry.

\subsubsection{Bi-polar vs. Single-Sided Ejections}
Our jet model assumes that each ejection event takes the form of two identical, oppositely directed plasmons. However, in principle our light curves could also be fit with a collection of single-sided ejections. These unpaired components could occur as a result of Doppler boosting of highly relativistic plasmons causing us to observe only the approaching component of an ejecta pair, or intrinsically unpaired ejecta.
Our simultaneous VLBA imaging may help distinguish between these two scenarios. We resolve both paired and (possibly\footnote{Given the rapid timescales of the ejections, multiple ejecta can become blended together in these images, making it difficult at times to conclusively identify and track individual components.}) unpaired ejecta components in our VLBA images, which could suggest that the emission in our light curves is produced by a combination of bi-polar and single-sided ejection events. Using these VLBA results to include stricter constraints within our model on ejecta numbers, type (single/bi-polar), and ejection times, is beyond the scope of this work, but will be considered in a future iteration of the model.

\subsection{Modelling Process and Best Fit Model}
Due to the large number of free parameters in our model,
we use a Bayesian approach for parameter estimation. In particular, we apply a Markov-Chain Monte Carlo algorithm (MCMC), implemented with the \textsc{emcee} package \citep{for2013},
to fit our light curves with our jet model. This package is a pure-Python implementation of Goodman \& Weare's Affine Invariant MCMC Ensemble Sampler \citep{goodwe10}, running a modified version of the commonly used Metropolis-Hastings Algorithm, whereby it simultaneously evolves an ensemble of ``walkers" through the parameter space. We use 500 walkers (10 $\times$ the number of dimensions in our model) for our MCMC runs. 
\renewcommand\tabcolsep{2.8pt}
\begin{table*}
\caption{V404 Cyg Jet Model Best Fit Parameters}\quad
\centering
\begin{tabular}{cccccccccc}
\hline\hline
\multicolumn{10}{c}{{\bf Compact Jet Parameters$^a$}}\\[0.15cm]
{${F_{0,\rm cj}}$ (mJy)}&{${\alpha}$}\\[0.25cm]\hline
$56.22^{+0.19}_{-0.21}$ & $-0.46^{+0.03}_{-0.03}$& \\[0.15cm]
&&&&&\\
\hline\hline
\multicolumn{10}{c}{{\bf Individual Jet Ejecta Parameters}}\\[0.15cm]
{Ejection \,} & 
{${t_{\rm ej}}\,$(HH:MM:SS.S)}&
   {${t_{\rm ej}}\,$(MJD)} &
   {${i\,}$(degrees)}& 
   {$\phi_{\rm obs}$(degrees)}&
   {${\tau_0\,}$}&
   {$p$$^b$}&
   {${F_0\,}$(mJy)}&
   {${\beta_{\rm b}\,}$ (v/c)} &
   {${\beta_{\rm exp}\,}$ (v/c)$^c$}\\[0.25cm]
   \hline
1 &$\mbox{\formattime{10}{23}{42}}.4 ^{+ 8.5 }_{- 8.0 }$ &$ 57195.4331 ^{+ 0.0001 }_{- 0.0001 }$ & $ 39.73 ^{+ 1.64 }_{- 1.57 }$ & $ 4.06 ^{+ 0.24 }_{- 0.22 }$ & $ 1.96 ^{+ 0.01 }_{- 0.01 }$ & $ 3.18 ^{+ 0.03 }_{- 0.03 }$ & $ 986.8 ^{+ 6.2 }_{- 5.4 }$ & $ 0.290 ^{+ 0.006 }_{- 0.006 }$& $ 0.014 ^{+ 0.001 }_{- 0.001 }$\\[0.25cm]
2 &$\mbox{\formattime{10}{36}{09}}.4 ^{+ 3.6}_{- 3.4 }$ &$ 57195.4418 ^{+ 0.0001 }_{- 0.0001 }$ & $ 58.80 ^{+ 1.37 }_{- 2.04 }$ & $ 9.86 ^{+ 0.73 }_{- 0.47 }$ & $ 2.60 ^{+ 0.01 }_{- 0.01 }$ & $ 5.69 ^{+ 0.01 }_{- 0.01 }$ & $ 1672.6 ^{+ 8.3 }_{- 9.3 }$ & $ 0.115 ^{+ 0.005 }_{- 0.007 }$& $ 0.017 ^{+ 0.002 }_{- 0.001 }$\\[0.25cm]
3 &$\mbox{\formattime{11}{21}{35}}.1 ^{+ 50.9 }_{- 46.6 }$ &$ 57195.4733 ^{+ 0.0006 }_{- 0.0005 }$ & $ 87.98 ^{+ 0.06 }_{- 0.07 }$ & $ 5.36 ^{+ 0.03 }_{- 0.03 }$ & $ 1.28 ^{+ 0.03 }_{- 0.03 }$ & $ 1.54 ^{+ 0.01 }_{- 0.01 }$ & $ 3909.1 ^{+ 95.1 }_{- 108.3 }$ & $ 0.574 ^{+ 0.011 }_{- 0.013 }$ &$ 0.066 ^{+ 0.002 }_{- 0.003 }$\\[0.25cm]
4 &$\mbox{\formattime{11}{28}{58}}.2 ^{+ 7.5 }_{- 7.3 }$ &$ 57195.4785 ^{+ 0.0001 }_{- 0.0001 }$ & $ 68.47 ^{+ 1.33 }_{- 1.42 }$ & $ 4.63 ^{+ 0.44 }_{- 0.39 }$ & $ 1.58 ^{+ 0.02 }_{- 0.02 }$ & $ 2.15 ^{+ 0.01 }_{- 0.01 }$ & $ 2050.1 ^{+ 8.5 }_{- 8.3 }$ & $ 0.392 ^{+ 0.006 }_{- 0.006 }$ &$ 0.032 ^{+ 0.003 }_{- 0.003 }$\\[0.25cm]
5$^d$ &$\mbox{\formattime{12}{30}{42}}.6 ^{+ 94.2 }_{- 99.6 }$ &$ 57195.5213 ^{+ 0.0011 }_{- 0.0012 }$ & $ 75.23 ^{+ 0.06 }_{- 0.05 }$ & $ 5.15 ^{+ 0.07 }_{- 0.07 }$ & $ 1.72 ^{+ 0.01 }_{- 0.01 }$ & $ 2.51 ^{+ 0.03 }_{- 0.03 }$ & $ 5496.2 ^{+ 186.8 }_{- 175.8 }$ & $ 0.861 ^{+ 0.003 }_{- 0.003 }$& $ 0.148 ^{+ 0.003 }_{- 0.003 }$\\[0.25cm]
6 &$\mbox{\formattime{12}{32}{47}}.6 ^{+ 87.4 }_{- 90.3 }$ &$ 57195.5228 ^{+ 0.0010 }_{- 0.0010 }$ & $ 85.51 ^{+ 0.08 }_{- 0.08 }$ & $ 6.06 ^{+ 0.03 }_{- 0.03 }$ & $ 1.71 ^{+ 0.01 }_{- 0.01 }$ & $ 2.48 ^{+ 0.02 }_{- 0.02 }$ & $ 2404.7 ^{+ 73.5 }_{- 65.0 }$ & $ 0.606 ^{+ 0.010 }_{- 0.010 }$ &$ 0.081 ^{+ 0.002 }_{- 0.002}$\\[0.25cm]
7 &$\mbox{\formattime{12}{39}{39}}.5 ^{+ 8.8 }_{- 9.4 }$ &$ 57195.5275 ^{+ 0.0001 }_{- 0.0001 }$ & $ 87.86 ^{+ 0.42 }_{- 0.28 }$ & $ 6.95 ^{+ 0.17 }_{- 0.17 }$ & $ 1.20 ^{+ 0.05 }_{- 0.03 }$ & $ 1.40 ^{+ 0.01 }_{- 0.01 }$ & $ 1756.4 ^{+ 11.8 }_{- 11.9 }$ & $ 0.186 ^{+ 0.005 }_{- 0.005 }$ &$ 0.023 ^{+ 0.001 }_{- 0.001 }$\\[0.25cm]
8 &$\mbox{\formattime{12}{42}{43}}.2 ^{+ 6.6 }_{- 6.9 }$ &$ 57195.5297 ^{+ 0.0001 }_{- 0.0001 }$ & $ 87.84 ^{+ 0.12 }_{- 0.20 }$ & $ 7.72 ^{+ 0.33 }_{- 0.21 }$ & $ 2.10 ^{+ 0.01 }_{- 0.01 }$ & $ 3.60 ^{+ 0.03 }_{- 0.04 }$ & $ 1491.9 ^{+ 13.8 }_{- 14.6 }$ & $ 0.085 ^{+ 0.002 }_{- 0.004 }$& $ 0.012 ^{+ 0.001 }_{- 0.001 }$\\[0.25cm]

\hline
\end{tabular}\\
\begin{flushleft}
{$^a$ {The emission from the underlying compact jet portion of our model is best fit by a single power-law rather than a broken power-law.}}\\
{$^b$ The index of the electron energy distribution, $p$, is not a fitted parameter but rather is solved for using values of $\tau_0$ and Equation 4.}\\
{$^c$The expansion velocity, $\beta_{\rm exp}$, is not a fitted parameter but rather is solved for using values of $\beta_{\rm b}$, $i$, $\phi_{\rm obs}$ and Equation 11.}\\
{$^d$We note that the receeding component for this ejection is not well constrained, as it is modelled primarily by the SMA data at later times when the VLA observations had stopped (see Figure~\ref{fig:modellcindv}). Therefore, the parameters for this ejection should be treated with caution.}\\
\end{flushleft}
\label{table:vdl}
\end{table*}
\renewcommand\tabcolsep{6pt}
\begin{figure*}
\begin{center}
 \includegraphics[width=1.2\columnwidth]{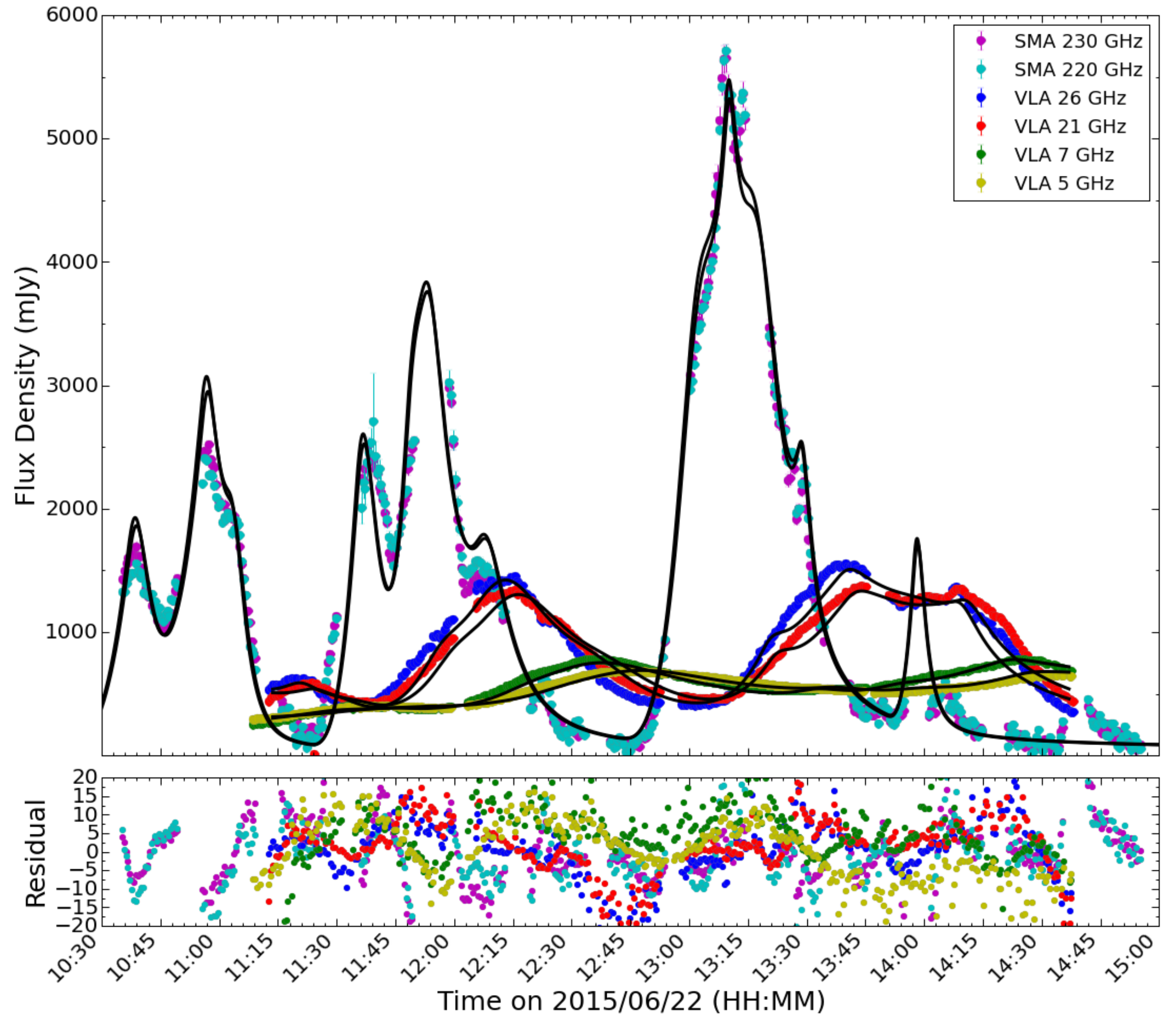}
 \caption{\label{fig:modellc}  Radio through sub-mm light curves of V404 Cyg on 2015 June 22. In the top panel we have overlaid the predicted best fit model at each frequency (black solid lines) on top of the light curves. The residuals are shown in the bottom panel, where, residual=(data-model)/(observational uncertainties). The JCMT 350 GHz data are not shown in this figure even though they are included in the fit. We do this for the sake of clarity in the figure, due to the small time lag between the JCMT and SMA data (see Figure~\ref{fig:modellcindv} for the JCMT 350 GHz light curve and model). With a total of 8 bi-polar ejection events, our model can reproduce the emission we observe from V404 Cyg at all of our sampled frequencies remarkably well.
}
\end{center}
\end{figure*}

\begin{figure*}
\begin{center}
 \includegraphics[width=1.2\columnwidth]{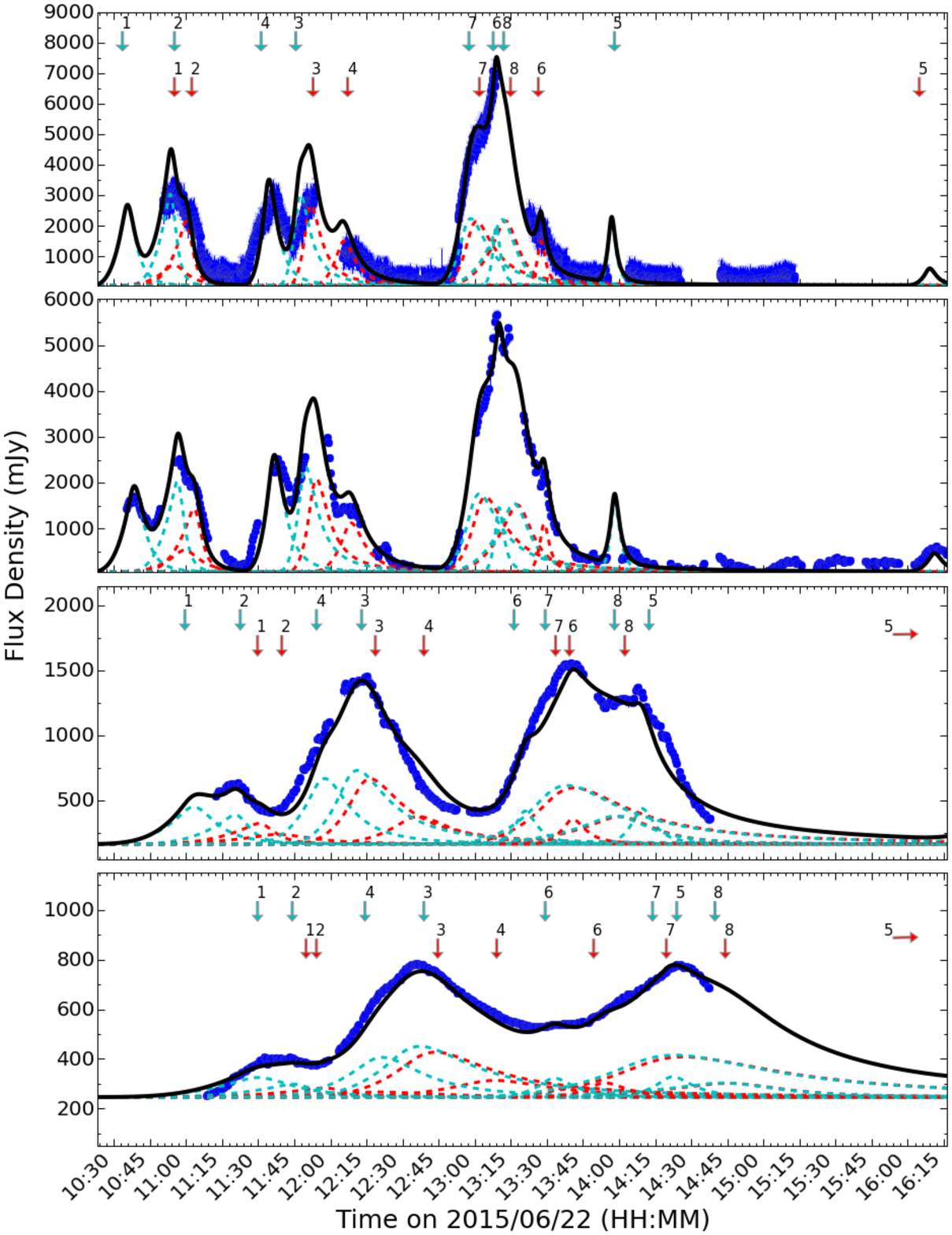}
 \caption{\label{fig:modellcindv}  V404 Cyg light curves at representative frequencies; 350 GHz (top), 230 GHz (2nd from top), 26 GHz (third from top), and 7 GHz (bottom). In all panels, the black solid line indicates the total model, and the dotted lines indicate the approaching (cyan) and receding (red) components of the individual ejection events. The arrows (cyan for approaching, red for receding) identify which flares correspond to which ejection number from Table~\ref{table:vdl}. Note that we do not attempt to model all of the sub-mm emission at times after the VLA observations had stopped.
}
\end{center}
\end{figure*}

Prior distributions used for all of our parameters are listed in Table~\ref{table:params}. We choose physically informative priors that reflect our knowledge of V404 Cyg (or commonly assumed values for BHXBs) where possible, and wide uninformative uniform priors when we have no pre-defined expectation for a specific parameter. 
For instance, the prior for the inclination angle is set as a truncated normal distribution, centered on 67 degrees (the measured inclination angle of the system), with boundaries of 0 and 90 degrees (allowed values of the inclination angle). On the other hand, the prior for the ejection time is simply a uniform distribution, sampling a wide range of possible times around our best initial guess.

Before running the MCMC,  the initial position of the walkers in the parameter space needs to be defined. As the performance of the \textsc{emcee} algorithm tends to benefit heavily from well defined initial conditions, we do an initial exploration of the parameter space using a harmony search global optimization algorithm\footnote{Implemented in the python package, \textsc{pyHarmonySearch}; https://github.com/gfairchild/pyHarmonySearch }.
This metaheuristic algorithm, that is similar to, but much more efficient than a brute force grid search method (which would not be computationally feasible in this case), yields a reasonable initial guess for our model, and we place our walkers in a tight ball around this initial guess in the parameter space.

As our jet model can predict emission at multiple frequencies, to reduce the degeneracy in our model, we choose to simultaneously fit all of our multi-frequency data sets, except for the JCMT 666 GHz data set, due to its sparser sampling and larger uncertainty in flux calibration (see the Appendix for details).
To do this, we use an iterative process whereby we start with our reference frequency data set, run the MCMC (the walkers are evolved over a series of steps, where the first 500 step ``burn in" period is not retained) until convergence is reached, and use the final position of the walkers for the first run as the initial guess for the next run of the MCMC, which will include increasingly more data sets in the fit.  To monitor the progress of the MCMC and ensure that correct sampling was occurring, we checked that the acceptance fraction stayed within the suggested bounds (between 0.25 and 0.75). Our criteria for convergence requires that the positions of the walkers are no longer significantly evolving. We determine whether this criteria is met by monitoring the chains of each of the walkers through the parameter space, and ensuring that, for each parameter, 
the intra-chain variance across samples is consistent with the inter-chain variance at a given sample.

Using the multi-dimensional posterior distribution output from the converged MCMC solution, we create one dimensional histograms for each parameter. The best fit result is taken as the median of these distributions, and the uncertainties are reported as the range between the median and the 15th percentile (-), and the 85th percentile and the median (+), corresponding approximately to $1\sigma$ errors. 
All of the best-fit parameters and their uncertainties are reported in Table~\ref{table:vdl}. Figure~\ref{fig:modellc} \& \ref{fig:modellcindv} show the best fit model overlaid on our multi-frequency light curves. Additionally, with our multi-dimensional posterior distribution we can explore the possible two-parameter correlations for our model, where a significant correlation between a pair of parameters can indicate a model degeneracy or a physical relationship between the parameters. In the Appendix section we show correlation plots (Figure~\ref{fig:corr}), along with the one-dimensional histograms, for pairs of parameters for which we find a correlation, and discuss the significance of such a correlation.

Within the Bayesian formalism, the uncertainties reported in Table~\ref{table:vdl} are purely statistical, and only represent
the credible ranges of the model parameters under the assumption that our model is correct. Given the residuals with 
respect to the optimal model (Figure~\ref{fig:modellc} bottom panel), the observations contain physical or observational effects not completely accounted for in our model. To factor in how well our chosen model represents the data, we estimated an additional systematic error for our parameters (displayed in Table~\ref{table:vdl_fix} of the Appendix). To do this we rerun our MCMC, starting from the best fit solution, with an extra variance parameter (effectively modelling all the physical/observational effects not included in our model) in our log probability for each frequency band. This variance is equivalent to the square of the mean absolute deviation of the residuals with respect to our optimal model at each frequency (difference between the best fit model and the data). The resulting uncertainties in the parameters after this extra MCMC run will reflect the full (statistical + systematic) uncertainties.
 
Our broad frequency coverage, in particular the high sub-mm frequencies, is crucial to the success of our modelling. Detailed substructure detected in the sub-mm bands can be used to separate out emission from different ejections, where their lower frequency counterparts are smoothed out and blended together.  As such, modelling the lower frequency emission would not be possible without the critical information the high frequency sub-mm emission provides and vice versa.

\section{Discussion of the Best Fit Model} 
Our jet model for V404 Cyg, with a total of 8 bi-polar ejection events on top of an underlying compact jet, is able to reproduce the emission in all of our observed frequency bands, matching the flux levels, time lags between frequencies, and the overall morphology remarkably well. 
With such a large sample of jet ejecta, we can probe the intrinsic ejecta properties, and the distribution of these properties between the different ejection events. In particular, our model characterizes the bulk speeds, peak fluxes, the electron population injected during each event, and the jet geometry, all of which we find can vary between events, with bulk speeds of $0.08<\beta_b<0.86$ c, peak fluxes of $986<S_0<5496$ mJy, electron energy distribution indices of $1.4<p<5.6$ (corresponding to $1.2<\tau_0<2.6$), and observed opening angles of $4.06<\phi_{\rm obs}<9.86$$^\circ$. 
In the following sections we discuss these ejecta parameters and what they can tell us about jet speeds, energetics, mass loss, and geometry. Additionally, we draw comparisons between the V404 Cyg ejection events and the jet oscillation events in GRS 1915+105, as well as other multi-wavelength observations of V404 Cyg.

\subsection{Jet Speeds}
The bulk speeds of jet ejecta measured in BHXBs\footnote{An important caveat when considering the value of the bulk Lorentz factor ($\Gamma$), estimated using proper motions of discrete jet ejecta, is that $\Gamma$ depends strongly on the assumed distance to the source \citep{fender03}. While the distance is well known for V404 Cyg, this is not the case for the majority of BHXBs, and as a result constraints on $\Gamma$ in these systems typically represent lower limits.} can vary from system to system (e.g., $\Gamma\sim 1$ in SS 433; \citealt{hjon81},  
$\Gamma\sim2$ in V4641 Sgr; \citealt{hj00b}), where some systems that are known to enter high luminosity states, like V404 Cyg, have been shown to launch jet ejecta with $\Gamma>2$ (e.g., GRO J1655-40; \citealt{hjr95}).
However, in V404 Cyg we find that the bulk speeds of our modelled ejecta are quite low, with bulk Lorentz factors of only $\Gamma\sim 1-1.3$ (excluding ejection 5; see footnote c in Table~\ref{table:vdl} for details).

Moreover, V404 Cyg shows bulk speeds that vary substantially between ejection events, on timescales as short as minutes to hours. There is some evidence in the literature that jet speeds can vary within a BHXB\footnote{There is also evidence of jet speeds varying in neutron star XBs, most notably, Sco X-1 \citep{fom01,foma01} and Cir X-1 \citep{tudose08}.} source. For example, \cite{blun05} find small variations in jet speed up to 10\% in SS 433, jet speeds have been reported to vary between outbursts of H1743-322 (\citealt{co05}; \citealt{millerj12}), and varying proper motions have been measured in GRS 1915+105 \citep{millerj07}. However, no other source has shown variations as large, or on as rapid timescales as V404 Cyg. 

Performing a Monte Carlo Spearman's rank correlation test, we find no correlation between jet speed and ejection time, where, for instance, the bulk speed of the ejections (i.e., $\beta_bc$) increased or decreased throughout our observation period. However, we find a potential correlation (Spearman coefficient of $0.83\pm0.07$ with a p-value of $0.01$) between bulk speed and peak flux of our modelled ejecta, where brighter ejecta tend to have higher speeds. This correlation is consistent with what was seen in H1743-322, where higher bulk ejecta speeds corresponded to higher radio luminosity measurements (\citealt{co05}; \citealt{millerj12}).

The factors that govern jet speed in BHXBs are not well understood, but our measurements of surprisingly slow speeds, which can vary between sequential jet ejection events, suggest that the properties of the compact object (i.e., black hole mass) or peak luminosity of the outburst are likely not the dominant factors that affect jet speed.

Additionally, given the varying bulk speeds between the ejection events, it is plausible that later, faster ejections could catch up to earlier, slower ejections. Such a collision between ejecta may result in a shock that could be as bright or even brighter than the initial ejections, and in turn produce a flaring profile that could mimic a new ejection event. While including ejecta collisions in our model is beyond the scope of this work, we briefly consider the possibility here by examining the bulk motion of all of the ejections. We find that a collision between ejection 3 and ejection 2 would occur at $\sim$ 11:30 (if they were ejected at the same PA), which is very close to the predicted ejection time of ejection 4. Moreover, ejection 4 has a bulk speed which is in between the bulk speeds of ejection 2 and 3, as we might expect for the bulk motion of the plasmon after such a collision. However, given that the jet appears to be rapidly precessing in V404 Cyg (Miller-Jones et al. 2017, in prep), ejection 2 and ejection 3 are launched at very different inclination angles, which would prevent such a collision from occurring.  Therefore, given the precessing jet, we find this collision scenario unlikely.

\subsection{Jet Energetics, Mass Loss, and Particle Acceleration}
In our model we assumed that the radiating electrons follow a power-law energy distribution.
The power-law index of this distribution, $p$, informs us about the population of accelerated electrons initially injected into each discrete jet component, where the value of this energy index is governed by the electron acceleration mechanism. 
Fermi acceleration by a single shock can produce values of $p\sim2-3$, which are typically found in XRBs \citep{bland87,bell87,mar01}. However, the energy index can take on a wider range of values under certain conditions, where for example, lower values of $p$ (which result in a more asymmetric flare profile) can be produced if the acceleration occurs in multiple shocks \citep{mel93}, or if the electrons carry away kinetic power from the shock \citep{drur81}, and higher values of $p$ could be produced in the presence of oblique shocks (although this case requires highly relativistic shocks to produce large $p$ values; \citealt{Ball92}). Distributions with values of $p>4$ are nearly indistinguishable from a thermal (Maxwellian) distribution, which in the shock acceleration paradigm, implies very little acceleration has occurred (a shock essentially takes an input thermal distribution of electrons and builds a power-law distribution up over time).
Magnetic reconnection in a relativistic plasma is another viable mechanism that can accelerate electrons into distributions with similar $p$ values to shock acceleration. In this case, smaller $p$ values can be produced in the case of a strongly magnetized plasma ($\sigma>10$; where $\sigma\equiv B^2/4\pi nmc^2$ represents the magnetization parameter), and larger $p$ values can be produced in a weakly magnetized plasma ($\sigma\sim1$; \citealt{sironi14,guo14,sironi16}). In either theory of particle acceleration, we would expect a link between the speed (for shock acceleration) or magnetization (for magnetic reconnection), and the energy index, $p$.

The energy indices of our modelled ejecta appear to vary between sequential ejection events, with $1.4<p<5.6$  (where we find no clear correlation between $p$ values and jet speed). These $p$ values could be produced by shock acceleration or magnetic reconnection (under the right conditions), although we would need to invoke different mechanisms to produce distributions in both the very low and very high $p$ regimes (e.g., 1.4 in ejection 7, and 5.7 in ejection 2), which is not entirely physical for a single source.  
Further, this significant range seen in our energy indices suggests that our model may not be capturing all of the complexities of these ejection events, where the more extreme values of the energy index could be mimicking the effect of physics that has not been included in our model. For instance, the vdL model assumes equipartition, but as the plasmons expand they must do work, which will result in some of the magnetic field dissipating into kinetic or thermal pressure, and in turn, the assumption of equipartition may break down. Simplifications in our model such as this could also explain the lack of expected correlation between our energy indices and the speed of the ejecta. A more rigorous treatment, which, for example, calculates the full synchrotron flux (and does not rely on the equipartition assumption), is beyond the scope of this work, but will be considered in future iterations of this model.

For synchrotron emitting clouds of plasma injected with our measured electron distributions, we estimate that the minimum energy\footnote{In our minimum energy calculations, we perform the full calculations outlined in \cite{long11}, where we integrate the electron energy distribution from $\nu_{\rm min}=150 \,{\rm MHz}$ to $\nu_{\rm max}=666 \,{\rm GHz}$. The minimum frequency represents the lowest radio detection with LOFAR on June 23 \& 24 \citep{broderickj15}, and the maximum frequency represents our highest frequency sub-mm detection. When we consider an electron-proton plasma, we assume the ratio of the energy in the protons over that of the electrons is $\frac{\epsilon_p}{\epsilon_e}=1$.} needed to produce each of our modelled ejection events range from  $5.0\times10^{35}<E_{\rm min}<3.5\times10^{38}\, {\rm erg}$, with minimum energy magnetic fields\footnote{We note that while these calculation assume equipartition, the system could be far from equipartition. In this case the magnetic field would not necessarily be equivalent to the minimum energy field, but rather could be either much higher or much lower.} on the order of a few Gauss ($1<B_{\rm min}<35$ G). Taking into account the duration of each event, these energies correspond to a mean power into each event ranging from $4.0\times10^{32}<P_{\rm min}<2.5\times10^{35}\, {\rm erg\, s}^{-1}$. Due to the slow bulk speeds of the ejecta, including the kinetic energy from the bulk motion (in an electron-positron plasma $E_{\rm KE}=(\Gamma-1)E_{\rm min}$) yields only slightly higher values of $4.1\times10^{32}<P_{\rm tot}<2.6\times10^{35}\, {\rm erg\, s}^{-1}$. The minimum energy and power released within each of our modelled ejection events is comparatively lower than estimated for other major ejection events in BHXBs ($E_{\rm min}\sim1\times10^{43}\,{\rm erg}$; e.g., \citealt{fengarmc99} and $P_{\rm tot}\sim10^{36}-10^{39}\,{\rm erg\, s}^{-1}$; e.g., \citealt{fend99, br7, curr14}). This difference is dominated by the difference in the estimated size of the emitting region, where the radii of our modelled ejecta are smaller than is normally estimated for major ejection events, and the low bulk speeds, which result in a much smaller kinetic energy contribution.
Considering that the flaring activity in V404 Cyg lasted $\sim 2$ weeks (and assuming our observations to be representative of this entire period), we estimate that the the total (minimum) energy (radiative + kinetic) released into jet ejections over the full flaring period is $\sim3.2\times10^{40}\, {\rm erg}$, which is more on par with typical energies estimated for major ejection events in BHXBs. This total energy is also comparable to that carried by the accretion disc wind ($\sim10^{41}\,{\rm erg}$)\footnote{A rough estimate of the energy lost in the accretion disc wind is equivalent to $E_{\rm wind}\sim(1/2)M_{\rm wind} v_{\rm wind}^2$. Using $M_{\rm wind}\sim 10^{-8}\,M_\odot$ and $v_{\rm wind}\sim1000\,{\rm km\,s}^{-1}$ \citep{mun16}, we estimate $E_{\rm wind}\sim10^{41}\,{\rm erg}$.}. 

{If we assume that the jet ejecta contain some baryonic content, in the form of one cold proton for every electron, we calculate that the mean power into each event (including the kinetic energy from bulk motion) ranges from $6.2\times10^{32}<P_{\rm tot}<3.8\times10^{35}\, {\rm erg\, s}^{-1}$.
In this baryonic case, we estimate a total mass lost through the jet in our observation period of $9.4\times10^{-13}\,M_{\odot}$ (corresponding to $7.2\times10^{-11}\,M_{\odot}$ over the 2 week flaring period).} To compare this jet mass loss to the mass accreted onto the black hole, we follow a procedure similar to \cite{mun16}, using simultaneous INTEGRAL X-ray observations (only including the harder ISGRI bands, ranging from 25-200 keV) to calculate the total energy radiated (integrated X-ray luminosity) during our observations. To do this we convert the count rate into flux in the 10--1000 keV band using a power law model with photon index $\Gamma_p\sim1-2$, and approximating the integral as a sum ($\int L_X dt\approx\sum_i L_i \delta t=\bar{L}\Delta T$, where $\bar{L}$ is the weighted mean, $\delta t$ is size of the time bins, and $\Delta T$ is the total observation time). Assuming an accretion efficiency of 0.1, we calculate a total mass accreted during our observations of $M_{\rm acc,BH}=3.4\times10^{-11}-7.8\times10^{-11}\,M_{\odot}$. Therefore, the mass lost in the jet is a small fraction of the total mass accreted, ${M_{\rm jet}}=(1-3)\times10^{-2}\, M_{\rm acc,BH}\,$, and much less than the mass estimated to be lost in the accretion disc wind ($\sim 1000 M_{\rm acc,BH}$; \citealt{mun16}).

\subsection{Jet Geometry and Ejecta Size Scale}
Measurements of jet geometry in BHXBs, in particular the observed opening angle, only exist for a handful of systems, where all but one are upper limits (e.g., see Table~1 in \citealt{millerj06}, as well as \citealt{yang10} and \citealt{rush17} for recent measurements in XTE J1752-223 and XTE J1908+094).
Our simultaneous light curve modelling technique allows us to directly derive the first measurements of the jet geometry in V404 Cyg, where we model observed jet opening angles of $4.06<\phi_{\rm obs}<9.86$$^\circ$. These measurements are consistent with the opening angle estimates for the other BHXB systems with constraints, where the majority show $\phi_{\rm obs}\lesssim10^\circ$. 

With the opening angles, we can estimate the level of confinement of the jets in V404 Cyg by solving for the intrinsic expansion speed (using Equation 11; see last column of Table~\ref{table:vdl}) of our modelled ejecta ($\beta_{\rm exp}c=\frac{c}{\sqrt{3}}$ indicates freely expanding components, where $\frac{c}{\sqrt{3}}$ represents the speed of sound in a relativistic gas). We find intrinsic expansion speeds of  $0.01<\beta_{\rm exp}<0.1$ c, indicating a highly confined jet in V404 Cyg. There are many possible mechanisms that could be responsible for confining the jets in V404 Cyg. 
In particular: the jet could be inertially confined \citep{ic92}, where the ram pressure of the strong accretion disc wind detected in V404 Cyg  \citep{mun16} could inhibit the jet ejecta expansion\footnote{Although, we note that if the confinement is external, this would suggest that a very large amount of pressure surrounds the ejections. If this is supplied solely by the ram pressure from an accretion disc wind, then the mass-loss rate (proportional to the velocity ratio of the ejections to the wind) would be unrealistically large (i.e., greater then the mass accretion rate).}; the jet could be magnetically confined by a toroidal magnetic field \citep{eich93}; the jet could contain cold protons, which may impede the jet ejecta expansion \citep{millerj06}; or a combination of these different mechanisms could be at work.

Further, as we alluded to in the previous section, the initial radii of the jet ejecta (i.e., the radii of the ejecta at the time the sub-mm emission peaks)  estimated by our model are noticeably smaller than those typically estimated for major ejection events in other BHXBs.
This is likely a result of the much slower expansion velocities ($\beta_{\rm exp}<<1$) we find for the V404 Cyg ejecta. 
In particular, we infer a range of initial radii for our ejecta ranging from $(0.6-1.3)\times 10^{12} \, {\rm cm}$.
These radii appear to remain similar (to within a factor of 2) between ejection events.

\begin{figure*}
\begin{center}
 \includegraphics[width=13cm, height=9cm]{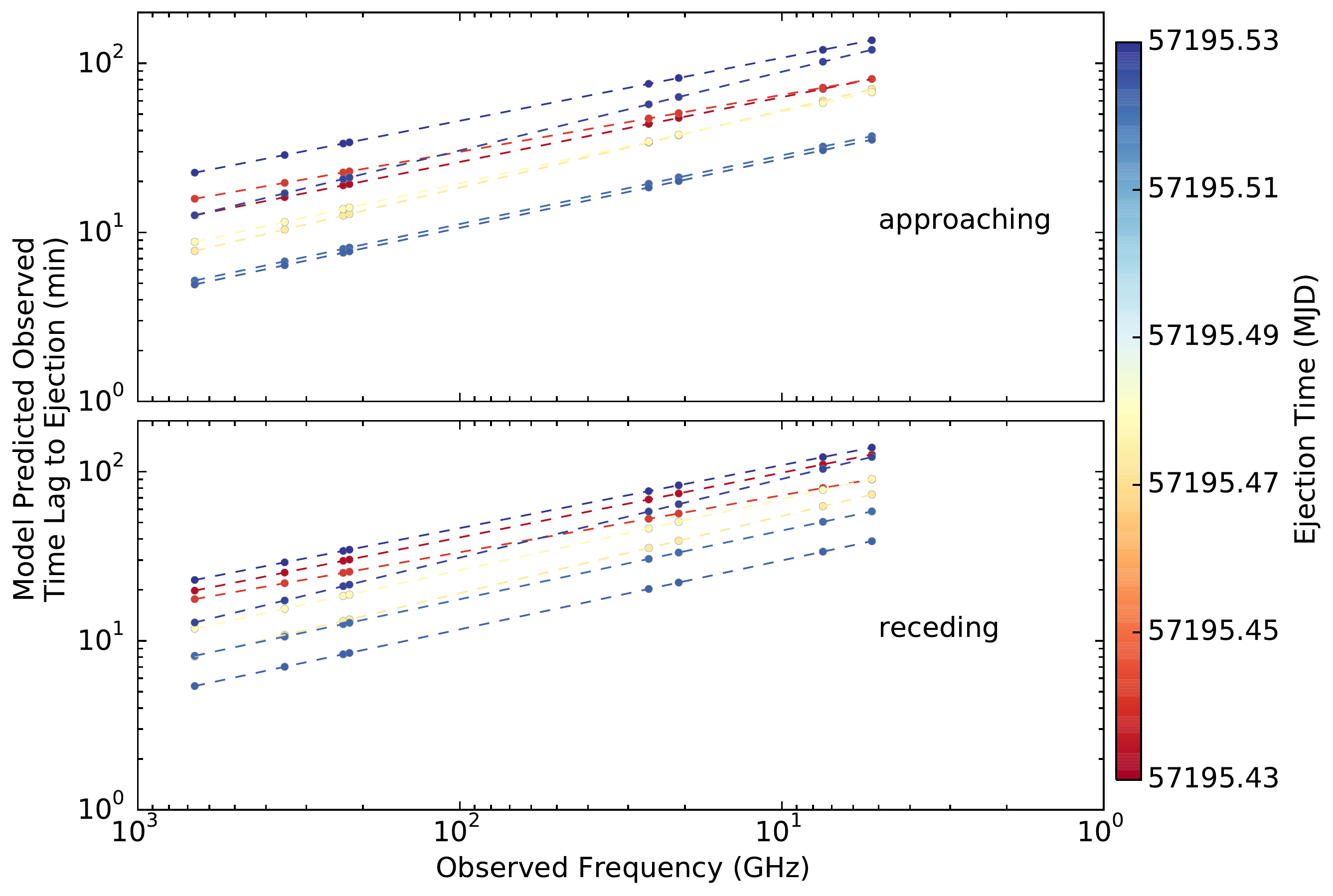}
 \caption{\label{fig:lags} Observed time lag, predicted by our model, between our sampled frequency bands and the time of ejection, for the approaching (top panel) and receding (bottom panel) components of each ejection event. The data points are coloured by ejection time, where the colour bar indicates the ejection time of the event in MJD.
}
\end{center}
\end{figure*}

\subsection{Underlying Compact Jet}
In addition to the jet ejecta component, we observe an extra constant flux component in our light curves, which varies with frequency.
Due to the shape of our estimated spectrum of this emission (see Figure~\ref{fig:blf}) and the strong compact core jet present throughout the span of our simultaneous VLBA imaging (Miller-Jones et al. 2017, in prep.), we interpret this extra flux term as emission from an underlying compact jet. We believe that this compact jet was switched on during the launching of the multiple discrete ejection events.
In our best fit model, this compact jet emission is characterized by a single power-law spectrum, with an optically thin spectral index of $\alpha=-0.46^{+0.03}_{-0.03}$. 

Our suggestion of a compact jet, that has not been fully quenched, is in agreement with the findings of \cite{san16}, who show that V404 Cyg never fully reached a soft accretion state (where we would likely expect strong quenching of the compact jet; e.g., \citealt{rush16}), but rather remained in either a harder intermediate or very high state during our observations.
Under this interpretation, based on our lowest radio frequency measurement, we place limits on the optically thick to thin jet spectral break frequency of $\nu_{\rm br}<5.25$ GHz, and flux at the spectral break of $S_{\rm br}>318$ mJy. However, we note that simultaneous VLITE observations \citep{kassimne15a} detect V404 Cyg at a total time-averaged flux density of $186\pm6$ mJy at 341 MHz. 
Given that our best fit model predicts a maximum jet ejecta flux component of $\sim 100$ mJy at 341 MHz, it is clear that the 341 MHz compact jet component cannot lie along the single power-law stated above. As such, the spectral break would therefore occur within the range of $0.341<\nu_{\rm br}<5.25$ GHz, which is significantly lower than previous estimates for V404 Cyg made during the hard accretion state ($\nu_{\rm br}=1.82\pm0.27\times10^5\, {\rm GHz}$; \citealt{rus12,rus13}). This evolution in the location of the spectral break is consistent with the pattern suggested by recent observations (e.g., \citealt{corb13,van13,rus14}) and MHD simulations \citep{pol14}, where, as the mass accretion rate increases during softer accretion states of BHXB outbursts (which usually occur at high luminosities; \citealt{kolj15}), the jet spectral break moves toward lower radio frequencies prior to the jet switching off (or at least fading below our detection limits). 

Up to now we have only considered the compact jet and the ejection events as separate entities. In the presence of explosive, energetic ejection events, we might expect a compact jet to be disrupted. In particular, as the ejecta collide with the pre-existing compact jet, a shock would likely develop, due to the difference in bulk speeds between the two. 
In this situation, if the compact jet rapidly re-establishes itself after being destroyed by ejecta (before the ejecta propagate far enough away from the black hole to be resolved), we would observe a compact core jet which appears to never shut off.
Therefore, we believe it is plausible that a compact jet is being repeatedly destroyed and re-established (on rapid timescales) following ejection events in V404 Cyg. Further, the emission from such a shock interaction could display an optically thin spectrum (similar to the interaction between the discrete ejecta and the surrounding ISM; e.g., \citealt{co04}), like the one we observe for our baseline emission component. 
Thus, while we interpreted the baseline emission in our light curves as originating only from a compact jet, emission from a possible interaction of the jet ejecta with this compact jet, and/or continuous lower-level, fainter jet ejecta that never get resolved, could also be contributing to the baseline flux level we observe.

Moreover, in our model we have assumed that the compact jet flux component is constant in time. However, as the accretion rate (and in turn the jet power) changes, the flux of a compact jet is expected to change as well \citep{rus14}. If we consider the erratic X-ray behaviour observed in the source (which presumably traces a rapidly changing accretion rate), it is plausible that the compact jet component could in fact be variable as well.
Exploring the possibility of a variable compact jet component in our model is left for future work.
 
\subsection{Ejecta Time Lags}
Our model predicts that the intrinsic time lag (in the source frame) between a certain frequency ($\nu$) and the reference frequency ($\nu_0$), is represented by,
\begin{equation}
t_{\nu-\nu_0,{\rm src}}=\left(\frac{R_0}{\beta_{\rm exp}}\right)\left\{\left(\frac{\nu_0}{\nu}\right)^{\frac{p+4}{4p+6}}-1\right\}
\end{equation}
where the observed time lag can be obtained through the transformation, $t_{\nu-\nu_0,{\rm obs}}=\frac{t_{\nu-\nu_0,{\rm src}}}{\delta_\mp}$.

Figure~\ref{fig:lags} shows the observed time lags, predicted by our model, between each frequency band and the ejection time, for the approaching (top panel) and receding (bottom panel) components. The time lags are clearly variable between different ejection events (e.g., $\sim10-30$ min between the ejection and our reference frequency, 230 GHz), which is a result of the varying ejecta properties (i.e., $\beta_{\rm exp}$, $p$, $R_0$).

Further, it is interesting to note that, for a different flaring event that occurred $\sim 2$ days after our data set, \cite{shahb16} measured a time lag of 2.0 hours \& 3.8 hours between the predicted ejection time (indicated by an r'-band polarization flare, which these authors suggest could be the signature of the launching of major jet ejection event) and the flare peaks at 16 GHz \& 5 GHz, respectively. These lags are slightly higher than predicted for the approaching components of our modelled ejection events, but appear to share a similar slope across frequencies.

\subsection{Comparison to GRS 1915+105} 
GRS 1915+105 is the only other BHXB in which a similar multi-frequency variability pattern to that seen in V404 Cyg has been reported. While flaring activity has been seen in other systems, it is often only detected in one frequency band (e.g., V4641 Sgr in optical; \citealt{uem04}), or the flares in question evolved over much longer (days rather than minutes/hours) timescales (e.g., 4U 1630-47 in radio/X-ray; \citealt{hje99}).
GRS 1915+105 has displayed some correlated radio, sub-mm, and IR flares (with lower frequency emission delayed from higher frequency emission), which repeated every $\sim20$ minutes for a $\geq10$ day period \citep{fenpoo00a}. While no discrete components were resolved with VLBI during the events, the similar rise and decay times of flares at different frequencies suggest that adiabatic energy losses, likely during the expansion of discrete components, played a key role in determining the flaring profiles of these events.
In fact, \cite{mir98} found that the timing of the radio emission during these events was consistent with synchrotron emission from adiabatically expanding plasma clouds, where each event required an energy input of $\sim10^{39}\,{\rm erg}$, and carried an estimated mass of $\sim10^{18}\,{\rm g}$. Many studies of these jet ejection events suggest that they occur as a result of instabilities causing the repeated ejection and refilling of the inner accretion disc or coronal flow (e.g., \citealt{bello97,nan01,vad01}).

\begin{figure*}
\begin{center}
 \includegraphics[width=2\columnwidth, height=21cm]{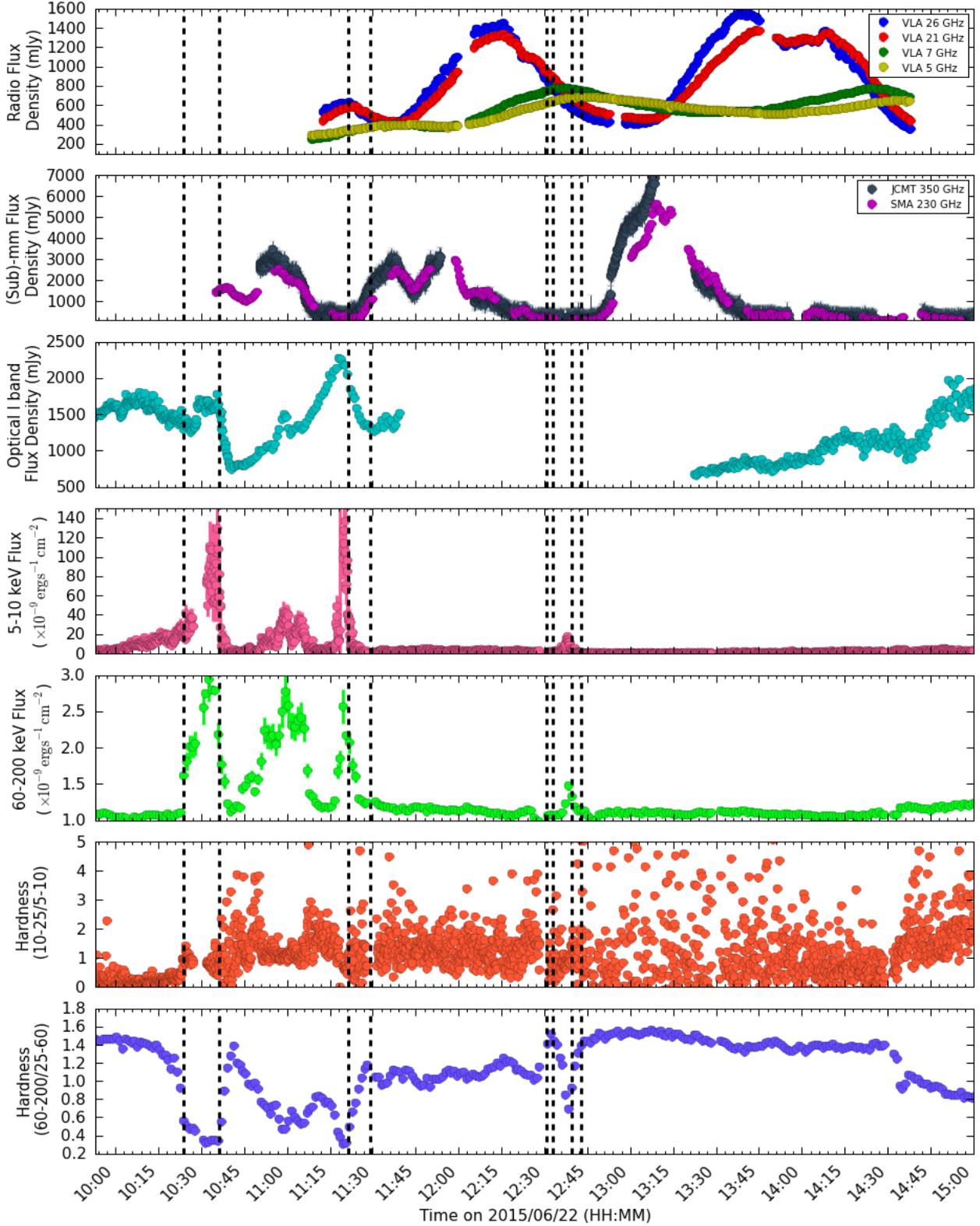}
 \caption{\label{fig:ejtimes_all}  The evolution of the emission properties of V404 Cyg on June 22. Top to bottom the panels represent radio light curves, (sub)-mm light curves, optical light curve \citep{kim16}, soft and hard X-ray light curves, and the 10-15/5-10 keV and 60-100/25-60 keV hardness ratios \citep{rod15aa}. Our modelled ejection times are shown by the dotted vertical lines, where the uncertainties on the ejection times are smaller than the line thickness.
}
\end{center}
\end{figure*}

In V404 Cyg our modelled ejection times appear to occur on a similar rapid timescale as seen in GRS 1915+105, where we observe groups of 2-4 ejections (separated by at most $\sim20$ minutes), followed by longer periods of up to $\sim 1$ hour between groups (see Figure~\ref{fig:ejtimes_all}). Each group of ejections seems to correspond to a large flaring event in the light curves.
Our estimates of the energetics and mass-loss of the V404 Cyg events (\S 5.2) are also similar to those estimated for the oscillation events in GRS 1915+105, where both are consistent with being smaller-scale analogues of major ejection events seen in other BHXBs. Further, \cite{naik01} suggested that multiple ejections in GRS 1915+105 could manifest as a single radio flare, similar to the ejection groupings we see in V404 Cyg. However, a noticeable difference in the timing of the V404 Cyg and GRS 1915+105 events is that the V404 Cyg events are not as quasi-periodic (i.e., they do not occur on as regular intervals) when compared to the GRS 1915+105 events, which occurred every $\sim20$ min \citep{fenpoo00a}. The absence of quasi-periodicity in the V404 Cyg events could indicate that the jet production process is not as stable in V404 Cyg as it was during the GRS 1915+105 events.

The similarity between the morphology, duration and energetics of the rapid ejection periods in V404 Cyg and GRS 1915+105 suggests that the events may have a common origin, possibly in the repeated ejection and refilling of some reservoir in the inner accretion flow.  
This hypothesis is consistent with the recent finding of \cite{rad16}, who report the non-detection of the disc component in the X-ray spectra following major radio flares in V404 Cyg. Although, given the large intrinsic absorption \citep{mott16b} seen in V404 Cyg during the outburst, it is conceivable that we may not have been able to detect the soft disc component, even if it was present. Nevertheless, as both V404 Cyg and GRS 1915+105 are long period systems, with large accretion discs, a key ingredient in fuelling rapid, repeated ejection events may be a large accretion disc (as suggested by \citealt{kim16,mun16}).

\subsection{Alternative Emission Models} 
Other than the vdL model, an alternative emission model that has been used to reproduce flaring light curves in XRBs is the shock-in-jet model \citep{mars85,bjorn00,turler00}. This analytical model, while traditionally favoured for extragalactic sources, has been successfully applied to flaring events in Cyg X--3 \citep{lin07,millerjtur09}, $\rm{ GRO\,\, J}1655-40$ \citep{stev03}, and GRS 1915+105 \citep{turl04}. The shock-in-jet model considers a shock wave travelling downstream in a jet flow as the source of each flare in the light curve. Each shock wave will evolve through three different phases; (1) Compton losses dominate, (2) synchrotron losses dominate, and (3) adiabatic losses dominate. The main differences between the shock-in-jet model and the vdL model are that the shock-in-jet model considers an initial phase where Compton losses dominate over adiabatic losses, all shock wave events are self-similar, and the electron energy scales differently when compared to the vdL model (shock-in-jet flow expands in 2D, $E\propto R^{-2/3}$;  vdL cloud expands in 3D, $E\propto R^{-1}$). These differences will result in a different flare profile between models, where for the same electron population (i.e., same $p$ value), the shock-in-jet model flares will show a much shallower decay, and the peak fluxes at frequencies that are initially optically thin (likely IR and above) will be smaller than predicted by the vdL model (which will over predict peak fluxes at these frequencies).

As our adapted vdL model is able to reproduce all our light curves (at 7 different frequencies) remarkably well, and simultaneous VLBA imaging  resolves multiple, discrete components (Miller-Jones et al. 2017, in prep), we favour the expanding plasmon model over the shock-in-jet model for the V404 Cyg events (although we can not rule out the shock-in-jet model).

However, for the GRS 1915+105 oscillation events, the emission has been shown to be consistent with both an expanding plasmon model (\citealt{mir98}; although we note that these authors only model a single flaring event, and did not include any relativistic/projection effects in their model) and a shock-in-jet model \citep{turl04}. If the GRS 1915+105 events are in fact a result of shock waves rather than expanding plasmons, this could explain the notable differences to the V404 Cyg events, namely the quasi-periodicity and the lack of VLBI resolved components\footnote{Although, we note that these GRS 1915+105 oscillation events were only observed with MERLIN \citep{fend99}, which does not have the resolution to see ejection events of a few mas in size (like those of V404 Cyg).}. Additionally, as \cite{turl04} point out, the shock-in-jet model is still consistent with the scenario that these oscillation events originated with instabilities in the inner accretion disc, as these instabilities could be the catalyst that leads to an increased injection rate of material at the base of the jet, and in turn a downstream shock wave.

\subsection{Connection to X-ray \& OIR}
If the jet ejection events in V404 Cyg are linked to processes occurring in the accretion flow, we might expect our predicted ejection times to correlate well with X-ray/OIR emission. For instance, in GRS 1915+105, IR and radio flares (which are presumably tracers of the ejection events) followed an X-ray peak and occurred during a period of spectral softening (dips in hard X-ray emission). However, the connection is not as clear in V404 Cyg. Figure~\ref{fig:ejtimes_all} displays our predicted ejection times on top of simultaneous X-ray\footnote{All X-ray data presented in this paper are taken from the INTEGRAL public data products available at http://www.isdc.unige.ch/integral/analysis\#QLAsources.} \citep{rod15aa} and OIR \citep{kim16} emission.

Flares in the OIR light curve appear to coincide with flares in the X-ray light curves. However, an unfortunate gap in the OIR coverage makes it difficult to confirm that such a pattern holds for the final X-ray flare. 
In terms of our modelled ejection times, we may be able to tentatively match groups of ejections with specific X-ray/OIR peaks, and possibly local dips in hardness (where the start/end of a steep gradient in hardness appears to correspond to ejections). But it is puzzling that the group which contains the largest number of ejections and produces the largest sub-mm flares appears to be connected to the X-ray flare with the smallest amplitude {(although, if an X-ray flare is indicative of a strong dissipative process, more energy dissipated in the X-ray implies less energy would be available to the jets, and vice versa).} Further, the second X-ray flare appears to have no jet ejecta counterpart.  

Given the extremely high intrinsic absorption during this time period \citep{mott16b}, it is entirely possible that the flaring in the X-ray light curves is not always dominated by intrinsic source variation, but rather dependent on how much of the inner accretion flow is obscured. This effect was seen in the 1989 outburst, where large changes in column density were determined to be the origin of some of the extreme X-ray variability observed \citep{oos97,z99}. Thus, even if the jet ejections are linked to processes in the accretion flow, we may not expect to see a clear correlation between our jet ejections and the X-ray/OIR emission. On the other hand, if the high absorption reduced the X-ray flux artificially, we would expect the high energy bands (60-200 keV) to be less affected than the lower energy bands (5-10 keV), which does not seem to be the case here. Therefore, the nature of the connection (if any) between our jet ejections and the X-ray/OIR emission is still not fully understood.

\subsection{The Critical Sub-mm Perspective}
Traditionally XRB jet studies have been dominated by radio frequency observations, such that there only exists a limited set of XRBs that have been observed at mm/sub-mm frequencies (e.g., \citealt{pared00,van13,rus13,fender01,tetarenkoa2015}). When considering time-resolved ($<1$ day cadence) mm/sub-mm observations this number decreases to two (i.e., GRS 1915+105; \citealt{poolf97}, Cygnus X-3; \citealt{baa86,fen95}). 
However, our work in this paper has clearly shown the vital importance of high time resolution mm/sub-mm data in XRB jet studies.
In particular, the mm/sub-mm bands can be used to isolate emission from different flaring events in the light curves, while the lower frequency counterparts of these events tend to be smoothed out and blended together.  As such, we find that radio frequency observations alone can often be misleading, especially in terms of identifying and pinpointing the timing of individual rapidly variable flaring events.
Including mm/sub-mm monitoring during future XRB outbursts will continue to add key insight to our understanding of jet behaviour.

\section{Summary}
In this paper we present the results of our simultaneous radio through sub-mm observations of the BHXB V404 Cyg during its June 2015 outburst, with the VLA, SMA and JCMT. Our comprehensive data set, taken on 2015 June 22 ($\sim 1$ week following the initial detection of the outburst), extends across 8 different frequency bands (5, 7, 21, 26, 220, 230, 350, and 666 GHz). 
Using custom procedures developed by our team, we created high time resolution light curves of V404 Cyg in all of our sampled frequency bands. In these light curves, we detect extraordinary multi-frequency variability in the form of multiple large amplitude flaring events, reaching Jy level fluxes. 

Based on the overall morphology, we postulate that our light curves were dominated by emission from a relativistic jet. To understand the source of the emission we constructed a detailed jet model for V404 Cyg. Our model is capable of reproducing emission from multiple, discrete, bi-polar plasma ejection events, which travel at bulk relativistic speeds (along a jet axis inclined to the line of sight), and evolve according to the van der Laan synchrotron bubble model \citep{vdl66}, on top of an underlying compact jet. 
Through implementing a Bayesian MCMC technique to simultaneously fit all of our multi-frequency light curves with our jet model, we find that a total of 8 bi-polar ejection events can reproduce the emission we observe in all of our sampled frequency bands. 

Using our best fit model to probe the intrinsic properties of the jet ejecta, we draw the following conclusions about the ejection events in V404 Cyg:
\begin{itemize}
\item The intrinsic properties of the jet ejecta (i.e., speeds, peak fluxes, electron energy distribution indices, opening angles) vary between different ejection events. This results in varying time lags between the flares produced by each ejection at different frequencies.
\item The ejecta require (minimum) energies on the order of $10^{35}-10^{38}\,{\rm erg}$. When taking into account the duration of each event, these energies correspond to a mean power into the ejection events of $10^{32}-10^{35}\,{\rm erg\,s}^{-1}$.
\item The ejecta carry very little mass ($\sim 1$\% $M_{\rm acc,BH}$), especially when compared to that carried by the other form of outflow detected in V404 Cyg, the accretion disc wind ($\sim 1000\, M_{\rm acc,BH}$). However, despite carrying much less mass, we estimate that the ejecta carry similar energy to that of the accretion disc winds.
\item We place the first constraints on jet geometry in V404 Cyg, where we find that V404 Cyg contains a highly confined jet, with observed opening angles of the ejecta ranging from $4.06-9.86^\circ$.  While we can not pin down the main jet confinement mechanism in V404 Cyg, it is possible that the ram pressure of the strong accretion disc wind detected in V404 Cyg \citep{mun16} could contribute to inhibiting the jet ejecta expansion, and thus be a key cause of the highly confined jet in this system.
\item The ejecta travel at reasonably slow bulk speeds, that can vary substantially between events, on timescales as short as minutes to hours ($\Gamma\sim 1-1.3$).
\item Brighter ejections tend to travel at faster bulk speeds.
\item Our modelled ejection events appear to occur in groups of 2-4 ejections (separated by at most $\sim 20$ minutes), followed by longer periods of up to $\sim 1$ hour between groups. 
\item The rapid timescale of the ejections is similar to the jet oscillation events observed in GRS 1915+105. Although the V404 Cyg events do not occur on as regular intervals as the GRS 1915+105 events, possibly suggesting the jet production process is not as stable in V404 Cyg.
\item We can tentatively match groups of ejections with peaks in simultaneous X-ray/OIR emission. However, the nature of the connection (if any) between our modelled ejection events and X-ray/OIR emission is still not completely clear.
\end{itemize}

Based on these conclusions, it appears as though the V404 Cyg ejection events are smaller-scale analogues of major ejection events, typically seen during the hard to soft accretion state transition in BHXBs.
Given the similarity between these rapid ejection events in V404 Cyg and those seen in GRS 1915+105, we 
postulate that the ejection events in both systems may have a common origin, in the repeated ejection and refilling of some reservoir in the inner accretion flow. This suggests that, in agreement with the findings of \cite{kim16} \& \cite{mun16}, the presence of a large accretion disc in both systems may be a key ingredient in producing these rare, rapid ejection events.

Overall, the success of our modelling has shown that, multiple expanding plasmons, on top of a compact jet, is a good match to the emission we observe from V404 Cyg in multiple frequency bands. However, it is also apparent from our results that some simplifications within our model may not fully capture all of the physics of these ejection events (e.g., assuming equipartition, assuming a constant flux from the compact jet), and future iterations of this model will work to address these assumptions and explore their effect on the ejecta properties.

In this work we have demonstrated that simultaneous multi-band photometry of outbursting BHXBs can provide a powerful probe of jet speed, structure, energetics, and geometry. Additionally, our analysis has revealed that the mm/sub-mm bands provide a critical new perspective on BHXB jets (especially in the time-domain) that can not be achieved with radio frequency observations alone. Future high time resolution, multi-band observations of more systems, including the mm/sub-mm bands, have the potential to provide invaluable insights into the underlying physics that drives jet behaviour, not only in BHXBs but across the black hole mass and power scale.

\section*{Acknowledgements}
We extend our sincere thanks to all of the NRAO, SMA, and JCMT staff involved in the scheduling and execution of these observations. Without their tireless hard work and constant support during this observing campaign we would never have obtained such an extraordinary data set.  We offer a special thanks to Iain Coulson for continuing to share his JCMT expertise. We thank M. Kimura et al. for sharing their OIR data. AJT thanks Eric Koch for many helpful discussions on MCMC implementation and cloud computing, and Patrick Crumley for his helpful comments and feedback on particle acceleration mechanisms. Also, many thanks to Christian Knigge for creating the V404 mailing list and Tom Marsh for creating the V404 observations website. Both of your efforts made it much easier to organize coordinated multi-frequency observations of this outburst. 
AJT is supported by an NSERC Post-Graduate Doctoral Scholarship (PGSD2-490318-2016). AJT, GRS, and EWR are supported by NSERC Discovery Grants. JCAMJ is the recipient of
an Australian Research Council Future Fellowship (FT140101082). SM acknowledges support from VICI grant Nr. 639.043.513/520, funded by the Netherlands Organisation for Scientific Research (NWO). TDR acknowledges support from the Netherlands Organisation for Scientific Research (NWO) Veni Fellowship, grant number 639.041.646. Cloud computing time on Amazon Web Services, used for the development and testing of our CASA timing scripts, was provided by the SKA/AWS Astro-Compute in the Cloud Program. Additionally, we acknowledge the use of Cybera Rapid Access Cloud Computing Resources, and Compute Canada WestGrid Cloud Services for this work.
 The National Radio Astronomy Observatory is a facility of the National Science Foundation operated under cooperative agreement by Associated Universities, Inc. The Sub-millimeter Array is a joint project between the Smithsonian Astrophysical Observatory and the Academia Sinica Institute of Astronomy and Astrophysics, and is funded by the Smithsonian Institution and the Academia Sinica.
 The James Clerk Maxwell Telescope is operated by the East Asian Observatory on behalf of The National Astronomical Observatory of Japan, Academia Sinica Institute of Astronomy and Astrophysics, the Korea Astronomy and Space Science Institute, the National Astronomical Observatories of China and the Chinese Academy of Sciences (Grant No. XDB09000000), with additional funding support from the Science and Technology Facilities Council of the United Kingdom and participating universities in the United Kingdom and Canada. 
 The authors also wish to recognize and acknowledge the very significant cultural role and reverence that the summit of Mauna Kea has always had within the indigenous Hawaiian community.  We are most fortunate to have the opportunity to conduct observations from this mountain.




\bibliography{ABrefList} 




\appendix
\section{Image Weighting Scheme}
{As we are imaging the source on very short timescales, the uv-coverage in each time-bin will be limited. While we do not need to worry about the lack of uv-coverage affecting the fidelity of the images, as the source is point-like at the VLA and SMA resolutions, the side-lobe levels may be a concern. In particular, if the amplitude is changing significantly in each time bin, this implies that we cannot deconvolve the side-lobes properly. 
As such, the choice of weighting scheme used while imaging could affect the quality of the images, and in turn the flux density measurements for each time bin.
While the side-lobe level is not much of a concern for the VLA, which has reasonably good instantaneous uv-coverage, the SMA is only an 8-element interferometer. In this case, imaging the source with a more uniform weighting scheme minimizes the side-lobe level, and could improve the quality of the images in each time bin. On the other hand, imaging with a natural weighting scheme would maximize sensitivity, leading to lower rms noise levels. After testing different weighting schemes we find that the choice of weighting had very little effect on the output SMA light curves, where any differences in the flux measurements in each time bin were well within the rms noise. We find that the natural weighting scheme led to lower rms noise and slightly higher dynamic range in the majority of the time bin images. Therefore, we opted to use natural weighting, as the side-lobe level/rms noise trade-off appears to be optimized for natural weighting.}

\section{Calibrator Light Curves}
\begin{figure*}
\begin{center}
 \includegraphics[width=18cm, height=7cm]{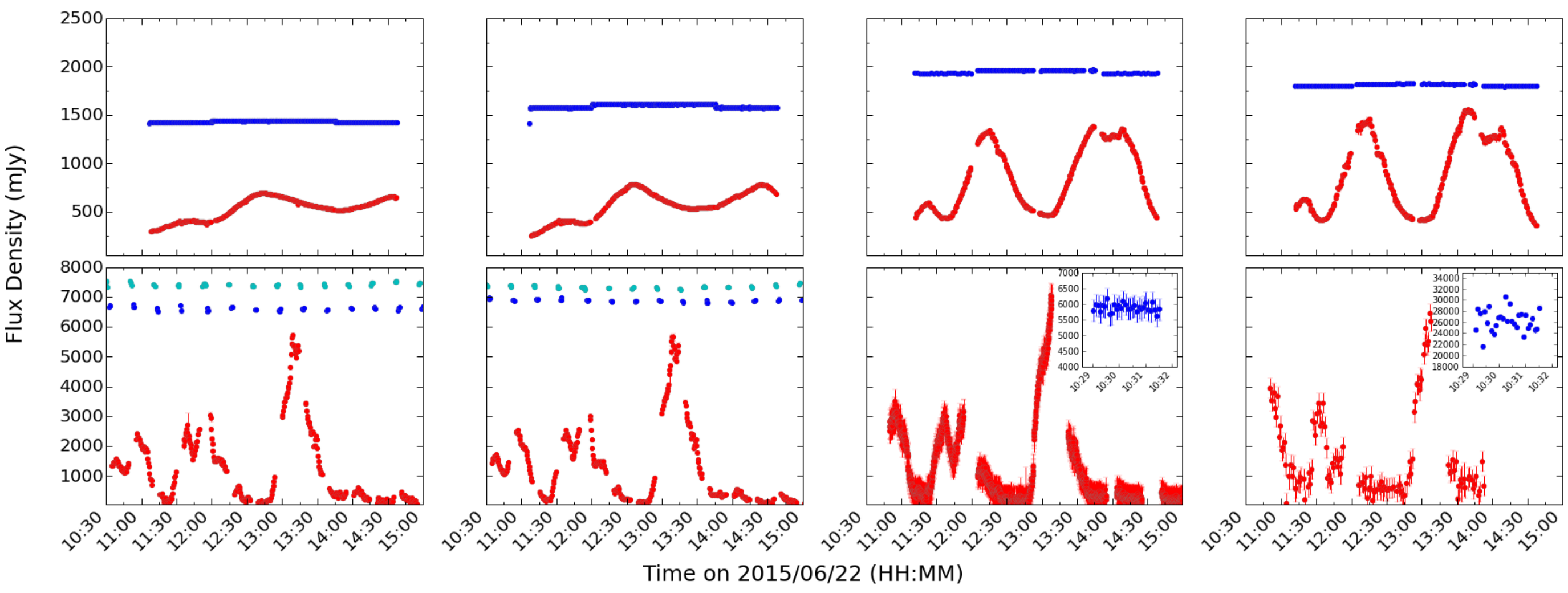}
 \caption{\label{fig:cals}   Multi-frequency light curves of V404 Cyg and our calibrator sources. \textit{(top)} Left to right the panels represent 5 GHz, 7GHz, 21 GHz, 26 GHz. \textit{(bottom)} Left to right the panels represent 220 GHz, 230 GHz, 350 GHz, and 666 GHz. In all panels, the calibrators are plotted in blue/cyan, while the V404 Cyg data are plotted in red. In the SMA data, the calibrator light curves are scaled up (5000 mJy added to total flux of the two calibrators) for clarity in the plot. In the JCMT data, the calibrator light curves are shown in inset panels as the calibrator scans were taken prior to the target scans.
}
\end{center}
\end{figure*}

Given the large flux variations we detected in our data of V404 Cyg, we wished to check the flux calibration accuracy of all of our observations on short time scales, and ensure that the variations we observed in V404 Cyg are dominated by intrinsic variations and not atmospheric or instrumental effects. Therefore, we ran our custom procedures to extract high time resolution measurements from our data (see \S3 for details) on all of our calibrator sources. Figure~\ref{fig:cals} displays target \& calibrator light curves at all frequencies.

We find that all of our interferometric calibrator sources and our JCMT 350 GHz calibrator display relatively constant fluxes throughout our observations, with any variations ($<5\%$/$<10\%$ of the average flux density at radio/(sub)-mm frequencies) being a very small fraction of the variations we see in V404 Cyg.
However, our JCMT 666 GHz calibrator scan shows noticeably larger scale variations ($\sim 30\%$ of its average flux level). While these larger variations are not unexpected at this high frequency, as the atmosphere is much more opaque, when combined with the fact that higher noise levels at this frequency prevent us from sampling timescales shorter than 60 seconds, we choose to not include the 666 GHz data set in our modelling (although see Appendix B below for a discussion of how well our best fit model agrees with the 666 GHz data).

Overall, based on these results, we are confident that the high time resolution light curves of V404 Cyg used in our modelling are an accurate representation of the rapidly changing intrinsic flux of the source.

\section{JCMT SCUBA-2 666 GHz Model Comparison}
\begin{figure*}
\begin{center}
 \includegraphics[width=13cm, height=7cm]{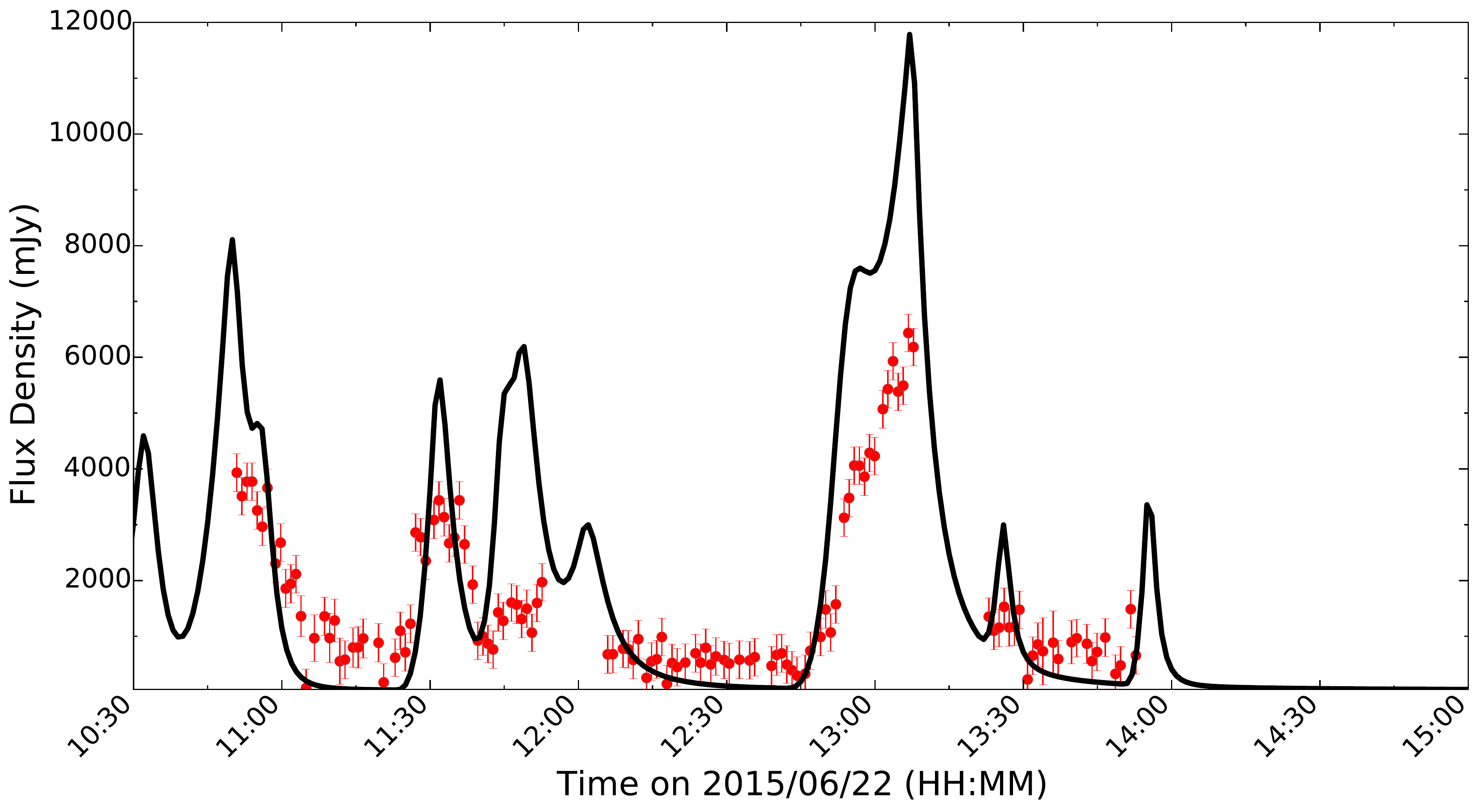}
 \caption{\label{fig:j666}   JCMT light curve of V404 Cyg in the 666 GHz band on 2015 June 22. The black solid line represents our predicted best fit model in the 666 GHz band. 
}
\end{center}
\end{figure*}

While we did not include the JCMT SCUBA-2 666 GHz data in our model fitting, it is still of interest to compare our best fit model prediction for the 666GHz band to the data (see Figure~\ref{fig:j666}). While our best fit model appears to match the timing of the flares in the 666 GHz data quite well, the model tends to over predict flux in some areas when compared to our data. It is possible that the deviations between the best fit model and the data are dominated by the higher flux calibration uncertainty in this band, especially when considering such short timescales. On the other hand, our model (and the vdL model) are only capable of predicting emission at frequencies which are initially self-absorbed (i.e. optically thick). Thus the deviations between the best fit model and the data could also suggest that the emission we observe from the jet ejecta in the 666 GHz band is initially optically thin.

\section{Systematic Errors}
As described in \S 4.3, we estimated additional uncertainties on our best fit parameters, to factor in how well our chosen model represents the data. Table~\ref{table:vdl_fix} displays these uncertainties (+ for upper confidence interval, $-$ for lower confidence interval) for each fitted parameter.

\renewcommand\tabcolsep{6pt}
\begin{table*}
\caption{Full (Statistical + Systematic) errors on V404 Cyg Jet Model Best Fit Parameters}\quad
\centering
\begin{tabular}{cccccccc}
\hline\hline
\multicolumn{8}{c}{{\bf Compact Jet Parameters}}\\[0.15cm]
{${F_{0,{\rm cj}}}$ (mJy)}&{${\alpha}$}\\[0.25cm]\hline
$ {+ 3.5 },{- 2.0 }$ & ${+0.05},{-0.06}$ \\[0.15cm]
&&&&&\\
\hline\hline
\multicolumn{8}{c}{{\bf Individual Jet Ejecta Parameters}}\\[0.15cm]
{Ejection \,} & 
{${t_{\rm ej}}\,$(sec)}&
   {${t_{\rm ej}}\,$(MJD)} &
   {${i\,}$(degrees)}& 
    {${\phi_{\rm obs}\,}$(degrees)}& 
   {${\tau_0\,}$}&
   {${F_0\,}$(mJy)}&
   {${\beta_{\rm b}\,}$ (v/c)} \\[0.25cm]
   \hline

1 &$ {+ 25.8 },{- 37.5 }$ &$ {+ 0.0003 },{- 0.0004 }$ & $ {+ 3.01 },{- 6.72 }$ & ${+ 1.00 },{- 0.34 }$ & ${+ 0.07 },{- 0.05 }$ &$ {+ 35.4 },{- 40.0 }$ & $ {+ 0.027 },{- 0.047 }$ \\[0.25cm]
2 &$ {+ 33.7 },{- 27.3 }$ &$ {+ 0.0004 },{- 0.0003 }$ & $ {+ 1.60 },{- 1.88 }$ & ${+ 0.43 },{- 0.41 }$ & ${+ 0.01 },{- 0.01 }$ &$ {+ 45.2 },{- 37.4 }$ & $ {+ 0.011 },{- 0.010 }$ \\[0.25cm]
3  &$ {+ 122.7 },{- 98.8 }$ &$ {+ 0.0014 },{- 0.0011 }$ & $ {+ 0.13 },{- 0.14 }$ & ${+ 0.16 },{- 0.13 }$ & ${+ 0.03 },{- 0.03 }$ & ${+ 325.4 },{- 342.5 }$ & $ {+ 0.017 },{- 0.024 }$ \\[0.25cm]
4 &$ {+ 28.3 },{- 27.9 }$ &$ {+ 0.0003 },{- 0.0003 }$ & $ {+ 2.45 },{- 3.54 }$ &$ {+ 0.83 },{- 0.51 }$ & ${+ 0.03 },{- 0.02 }$ &$ {+ 59.5 },{- 87.1 }$ & $ {+ 0.016 },{- 0.019 }$ \\[0.25cm]
5  &$ {+ 331.5 },{- 388.6 }$ &$ {+ 0.0038 },{- 0.0045 }$ & $ {+ 0.26 },{- 0.25 }$ & ${+ 0.35 },{- 0.29 }$ &$ {+ 0.13 },{- 0.11 }$ & ${+ 389.0 },{- 280.3 }$ & $ {+ 0.013 },{- 0.013 }$ \\[0.25cm]
6  &$ {+ 263.8 },{- 378.5 }$ &$ {+ 0.0031 },{- 0.0044 }$ & $ {+ 0.20 },{- 0.18 }$ & ${+ 0.73 },{- 0.92 }$ &$ {+ 0.14 },{- 0.11 }$ &$ {+ 483.8 },{- 291.9 }$ & $ {+ 0.036 },{- 0.026 }$ \\[0.25cm]
7  &$ {+ 27.4 },{- 26.7 }$ &$ {+ 0.0003 },{- 0.0003 }$ & $ {+ 1.22 },{- 0.72 }$ &$ {+ 0.29 },{- 0.23 }$ & ${+ 0.01 },{- 0.01 }$ & ${+ 51.1 },{- 51.6 }$ & $ {+ 0.009 },{- 0.010 }$ \\[0.25cm]
8  &$ {+ 45.9 },{- 55.5 }$ &$ {+ 0.0005 },{- 0.0006 }$ & $ {+ 0.48 },{- 0.96 }$ & ${+ 0.62 },{- 0.90 }$ &$ {+ 0.05 },{- 0.05 }$ & ${+ 85.2 },{- 118.0 }$ & $ {+ 0.007 },{- 0.006 }$ \\[0.25cm]

\hline
\end{tabular}
\label{table:vdl_fix}
\end{table*}
\renewcommand\tabcolsep{6pt}

\section{Two-Parameter Correlations}
\begin{landscape}
\begin{figure}
\begin{center}
 \includegraphics[width=24cm, height=14cm]{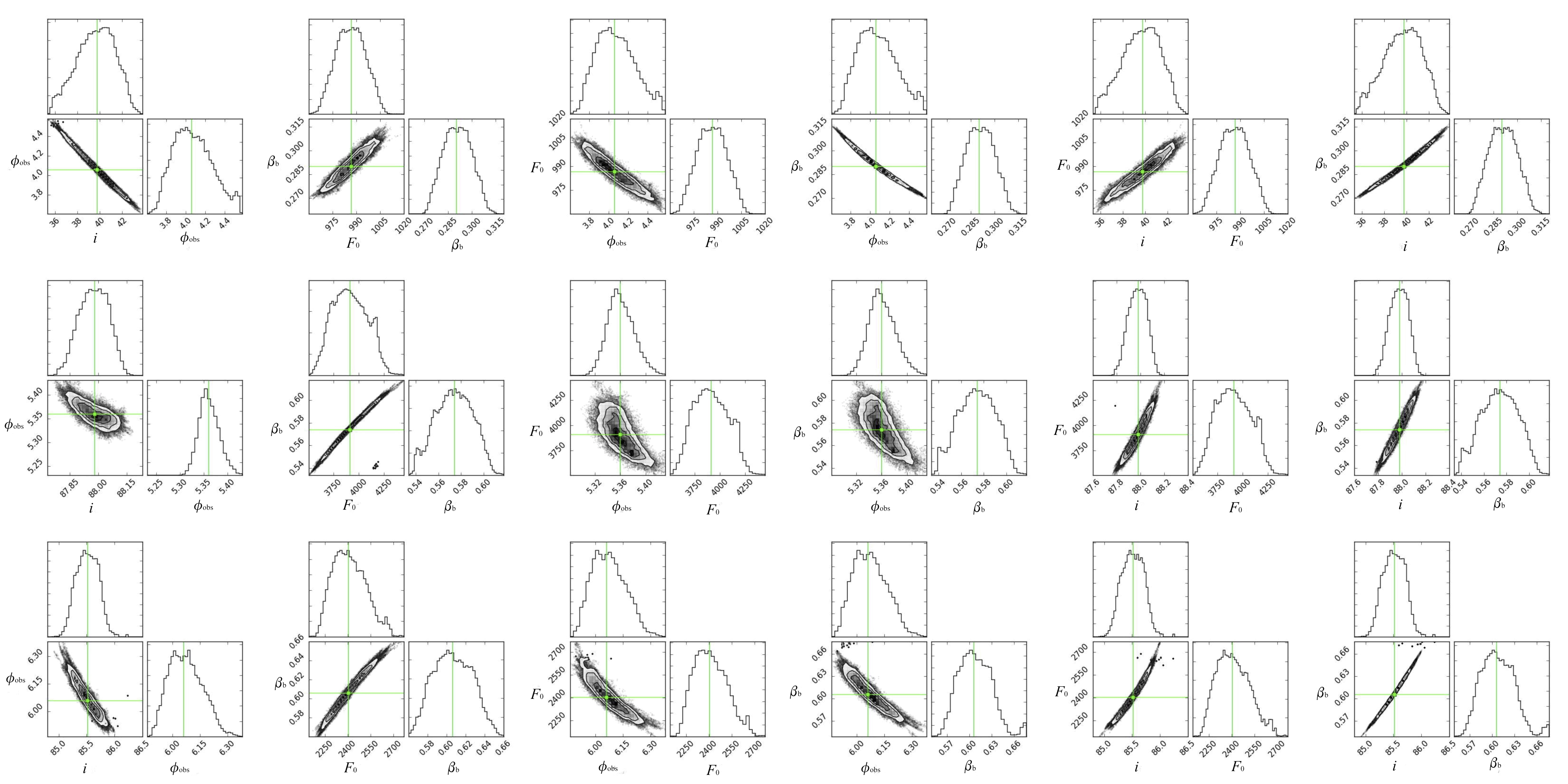}
 \caption{\label{fig:corr}  Two-parameter correlation plots for pairs of parameters in our model that show a significant relationship. While all the ejection events display the correlations shown, for clarity we only display the correlation plots for three events, which sample a range of $i$, $\phi_{\rm obs}$, $F_0$, and $\beta_b$ values, and are spread out in time throughout our observations; ejection 1 (top row), ejection 3 (middle row), ejection 6 (bottom row). The histograms represent the one dimensional posterior distributions of the parameters, and the green lines/squares indicate the best fit value of the parameters.
}
\end{center}
\end{figure}
\end{landscape}

With the multi-dimensional posterior distribution output from our MCMC runs, we explored possible two-parameter correlations for our model. A significant correlation between a pair of parameters, that is common to all of the ejecta, could indicate a model degeneracy or a physical relationship between the two parameters. Out of the possible two-parameter pairs, we find interesting correlations involving the $i$, $\phi_{\rm obs}$, $F_0$, and $\beta_b$ parameters. Figure~\ref{fig:corr} displays the correlation plots, along with the one-dimensional histograms of the parameters\footnote{We make use of the \textsc{corner} python module to make these correlation plots; https://github.com/dfm/corner.py}.
The correlation between $i$ and $\phi_{\rm obs}$ (first column) indicates a known degeneracy in the vdL model. The correlation between $F_0$ and $\beta_b$ (second column) likely indicates a physical relationship between the parameters, where faster ejecta tend to have brighter fluxes. We find the same relationship when we look at the distribution of bulk speeds and fluxes across all the ejecta, and this relationship has been seen in other sources (see \S 5.1 for details). The final four correlations (columns 3 through 6) seem to indicate a degeneracy between all four parameters (or at least a sub-set of them), where different combinations of the four parameters could potentially produce similar flaring profiles.

\bsp	
\label{lastpage}
\end{document}